\documentclass{aa}

\usepackage{graphicx}
\usepackage{txfonts}
\usepackage{xcolor}
\begin{document}

\title{CMASS galaxy sample and the ontological status of \\ the cosmological principle}
\subtitle{}

\author{Yigon Kim\inst{1}, Chan-Gyung Park\inst{2}\fnmsep\thanks{Corresponding author}, Hyerim Noh\inst{3} and Jai-chan Hwang\inst{1,4}   }

\institute{Department of Astronomy and Atmospheric Sciences, Kyungpook National University, Daegu 41566, Republic of Korea \\
              \email{yigon@outlook.com}
              \and
              Division of Science Education and Institute of Science Education, Jeonbuk National University, Jeonju 54896, Republic of Korea \\
              \email{park.chan.gyung@gmail.com}
              \and
              Theoretical Astrophysics Group, Korea Astronomy and Space Science Institute, Daejeon 34055, Republic of Korea
%              \email{hr@kasi.re.kr}
              \and
              Center for Theoretical Physics of the Universe, Institute for Basic Science (IBS), Daejeon 34126, Republic of Korea
%              \email{jchan@knu.ac.kr}
          }

%\date{Received September 15, 2021; accepted March 16, 1997}

% \abstract{}{}{}{}{}
% 5 {} token are mandatory

  \abstract
   {The cosmological principle (CP), {\it assuming} spatially homogeneous and isotropic background geometry in the cosmological scale, is a fundamental assumption in modern cosmology. Recent observations of the galaxy redshift survey provide relevant data to confront the principle with observation. Several previous studies claim that the homogeneity scale is reached at a radius around $70~h^{-1}\textrm{Mpc}$. However, the same observation shows a dramatic visual structure, SDSS Great Wall, which extends $300~h^{-1}\textrm{Mpc}$ in linear dimension.}
   {We present a homogeneity test for the matter distribution using the BOSS DR12 CMASS galaxy sample and clarify the ontological status of the CP.}
   {As a homogeneity criterion, we compare the observed data with similarly constructed random distributions using the number count in the truncated cones method. Comparisons are also made with three theoretical results using the same method: (i) the dark matter halo mock catalogs from the $N$-body simulation, (ii) the log-normal distributions derived from the theoretical matter power spectrum, and (iii) direct estimation from the theoretical power spectrum. }
   {We show that the observed distribution is statistically {\it impossible} as a random distribution up to $300~h^{-1}\textrm{Mpc}$ in radius, which is around the largest statistically available scale. However, comparisons with the three theoretical results show that the observed distribution is {\it consistent} with these theoretically derived results based on the CP.}
   {We show that the observed galaxy distribution (light) and the simulated dark matter distribution (matter) are quite inhomogeneous even on a large scale. Here, we clarify that there is no inconsistency surrounding the ontological status of the CP in cosmology. In practice, the CP is applied to the {\it metric} and the metric fluctuation is extremely small in {\it all} cosmological scales. This allows the CP to be valid as the averaged background. The matter fluctuation, however, is {\it decoupled} from the small nature of metric fluctuation in the subhorizon scale. What is directly related to the matter in Einstein's gravity is the curvature, a quadratic derivative of the metric.}


\titlerunning{CMASS galaxy sample and the ontological status of the cosmological principle}
\authorrunning{Y. Kim et al.}

\maketitle

%%%%%%%%%%%%%%%%%% BODY OF PAPER %%%%%%%%%%%%%%%%%%

\section{Introduction}

Modern cosmology explains diverse cosmological observations using a relatively simple model based on Einstein's gravity modified by the cosmological constant. The simple aspect of the model is not referring to Einstein's gravity (which is a set of highly nonlinear partial differential equations) nor to the enigmatic cosmological constant. A decisive simplification to make the problem mathematically tractable was made by \citet{einstein1917a}, and the assumption was later named as ``Einstein's Cosmological Principle'' \citep{milne1934,mccreamilne1934}. The assumption made in the 1917 paper was a broad-brush wish ``If we are concerned with the structure only on a large scale, we may represent the matter to ourselves as being uniformly distributed over enormous spaces.'' Einstein himself regarded it as too simplistic, mentioning in a letter to de Sitter in the same year \citep{einstein1917b} ``From the standpoint of astronomy, of course, I have erected but a lofty castle in the air.'' Concerning the validity of the assumption and its potential testability, he was somewhat pessimistic, mentioning in the same letter, ``Whether the model I formed for myself corresponds to reality is another question, about which we shall probably never gain information.''

Over the last 100 years, the cosmology Einstein has initiated with his Cosmological Principle eventually (because the actual history might have been more turbulent) turns out to be remarkably successful. It allows an easy theoretical treatment and leads to the eventual success of the theory matching with observations. The Principle now dominates the modern physical cosmology without his name attached. The Cosmological Principle (CP) in modern cosmology can be rephrased in a more precise term as follows: the universe is assumed to be homogeneous and isotropic on a sufficiently large scale. Our question to be addressed in this work, despite all the stunning successes in cosmology in recent years, concerns: what is the sufficient scale? And although Einstein mentioned the {\it matter} in the above quote, what is the CP referring to in cosmology: in matter or in geometry? This subtle point is vital in our argument, and even geometry needs later specification: in metric or in curvature? The matter is tied with geometry (curvature) in Einstein's gravity, but not so tightly concerning deviations from the CP (in the metric), especially on a small scale.

To clarify the case, let us introduce a term referring to the scale beyond which the CP is valid, the Homogeneity Scale (HS), in measurable quantities like the matter or the light. The HS depends on the definition of the homogeneity the cosmology needs and also on the subject the CP is applied. The homogeneity in theoretical cosmology is, in fact, a mathematical homogeneity in both matter and geometry, which is obviously an idealization {\it not} existing in nature. The CP represented in the Robertson-Walker metric first adopted by Friedmann is based on the mathematical homogeneity on spatial geometry \citep{friedmann1922,friedmann1924}. It demands the same homogeneity in the matter as well. Thus, the HS we wish to find out in the universe must be something different from this idealized mathematical definition (i.e., absolute homogeneity). What is it, then? Is the CP confirmed in observation? Where is the evidence? There is no doubt about the success of modern cosmology. But, what is the nature of the CP in modern cosmology? We will investigate these and other related questions in this work.

This paper has two main parts. In Section \ref{sec:homogeneity} we explain the status of the cosmic homogeneity and the nature of the CP in modern cosmology. In Section \ref{sec:test} we search for the HS in matter and light using the BOSS CMASS sample of galaxies by comparing with the random and a couple of mock catalogs based on numerical simulation and the power spectrum. We also show a visual presentation of the cosmic landscape, revealing significant inhomogeneity even on the largest statistically available scale. Section \ref{sec:discussion} is a discussion of future expectations and remaining issues.

\section{Cosmic homogeneity}
\label{sec:homogeneity}

Before we probe these questions further, let us examine the observational situations concerning the CP.

%%%%%%%%%%%%%%%%%%%%%%%%%%%%%%%%%%%%%%%%%%%%%%%%%%%%%%%%%%%%%%%
%
%
%
%%%%%%%%%%%%%%%%%%%%%%%%%%%%%%%%%%%%%%%%%%%%%%%%%%%%%%%%%%%%%%%
\subsection{The CP cannot be found in light or matter}

The cosmic microwave background radiation (CMB) shows a remarkable level of isotropy in the temperature distribution displayed on the celestial sphere centered around us: one-thousandth level in the dipole and $10^{-5}$ level in higher multiple anisotropies are detected \citep{cobe1992,wmap2003,planck2013}. {\it If} we disregard the dipole as the one caused by our motion, the remaining level of isotropy is indeed extraordinary. The isotropy of the CMB does not imply the homogeneity, however, of the matter distribution in the three-dimensional space enclosed by the recombination (last-scattering) surface of the CMB. The near-isotropy of the CMB implies near-homogeneity of the three-dimensional source distribution at the recombination epoch. But as the propagation of CMB light is affected by the spacetime geometry, the near-isotropy only implies near-homogeneity of the metric (geometry) fluctuation, which is {\it different} from the matter fluctuation. Moreover, the isotropic CMB does not exclude the strong inhomogeneity along the radial direction.

The set of angular locations of galaxies (or galaxy like objects) and their redshifts provides more direct data for homogeneity test on the cosmic scale. Different redshift implies different times in the past. Conversion of redshift into distance requires a cosmological model that already assumes the homogeneity in modern cosmology. Still, considering the success of the current cosmological model, we can proceed with the homogeneity test as a trial case. In this way, the success will be a {\it consistency check} of the CP, not proof. Three different samples are available for the test: the SDSS Main galaxies with $63,163$ objects nearby  ($z<0.2$), the Luminous Red Galaxies (LRG) sample with $105,831$ objects in the medium distance, and the CMASS sample with $931,517$ objects far away (in the past) \citep{eisenstein2001,abazajian2009,kazin2010,eisenstein2011,anderson2012,alam2015}. The SDSS Main Galaxy Sample shows a well-known panoramic structure with a colossal  $300~h^{-1}\textrm{Mpc}$ in linear scale, the Sloan Great Wall \citep{gott2005}. The LRG sample looks more homogeneous, and several studies were made concerning the homogeneity issue.

\citet{hogg2005}, using half of the final LRG objects, suggested  $70~h^{-1}\textrm{Mpc}$ in radius as the HS; there is even a consensus on this HS in the literature \citep{yadav2005,bagla2008,sarkar2009,yadav2010,scrimgeour2012,laurent2016,netlis2017}; see, however, \citet{labini2009,labini2010,labini2011} for a contrary view. Whereas \citet{cgpark2017}, using the full LRG objects, found {\it no} HS up to $300~h^{-1}\textrm{Mpc}$ in radius, which is around the largest statistically available scale. Both papers used the behavior of number counts within spheres (centered at the objects) as the radius of the sphere increases; while Hogg et al.\ used the centers of objects in the three-dimensional space, Park et al.\ also used the centers of objects in a narrowly chosen redshift range (thus losing the number of centers while minimizing mixed objects at different redshifts).
Different {\it criteria} for the homogeneity cause the different conclusions. Hogg et al.\ used a tendency of approaching the homogeneous distribution of the average number count with (in our opinion) an {\it ad hoc} cut-off criterion. Park et al.\ compared the galaxy counts with the random distribution (for both average and dispersion) as the homogeneity criterion and showed that the LRG distribution is statistically impossible to be achieved by the random distribution even up to the largest scale available, which is around $300~h^{-1}\textrm{Mpc}$ in radius. When one says distribution is homogeneous, what else can we compare besides the random Poisson one? Park et al.\ further compared the observation with the simulation results and showed their consistency; thus, the simulation based on the CP is similarly {\it not} homogeneous.

In this work, we extend the study of \citet{cgpark2017} now to the CMASS sample with more objects and far away (in both space and time). Our answer remains the same: {\it the CP is not in the sky}. The observed CMASS distribution (in both average and dispersion) is statistically {\it impossible} (the high $\sigma$'s appearing are staggering!) to be achieved by the random distribution even up to the largest scale available, which is still around $300~h^{-1}\textrm{Mpc}$ in radius, see Section \ref{sec:result}. This implies that the observed universe is far more clustered than the random distribution in {\it all} currently available observation scales. Thus, the CP is not in the {\it light} (which may or may not represent the matter).

We also compare the CMASS sample with mock galaxies generated from the $N$-body simulation and linear matter power spectrum.
Both mock data are based on the theory with the CP in the background. As we will show below, both data are consistent with the CMASS data and are far from homogeneity compared with the random data, see Section \ref{sec:result}. These comparisons confirm why the current cosmological model is successful, and at the same time, show that the CP cannot be found in these theoretical data, and nor is it demanded by the theory. The level of inhomogeneity in the CMASS sample is visualized in Fig.\ \ref{fig:CMASS-2D}. As both the $N$-body simulation and the linear matter power spectrum used to generate mock galaxies are based on the cold dark matter, our comparisons reveal that the CP is not in the {\it matters}, neither. Therefore, not in Nature.

%%%%%%%%%%%%%%%%%%%%%%%%%%%%%%%%%%%%%%%%%%%%%%%%%%%%%%%%%%%%%%%
%
%
%
%%%%%%%%%%%%%%%%%%%%%%%%%%%%%%%%%%%%%%%%%%%%%%%%%%%%%%%%%%%%%%%
\subsection{Resolution}

What does this seemingly contradicting situation imply? That is, the theoretical cosmology based on the CP is successful, while the CP cannot be found in the observed Universe.

The answer lies in distinguishing the CP theoretically applied to the {\it metric} (geometry) from the attempt to find it in the distribution of {\it matter} which includes luminous objects.
When we say geometry, it is crucial to distinguish between the metric and the curvature. The curvature is the second derivatives of the metric, and the spatial curvature is the second spatial derivatives of the metric. The point is that whereas the curvature is directly related to the matter in Einstein's equation, the CP is imposed on the metric. In this way, even with huge matter inhomogeneity, the metric (a double integration of the curvature inhomogeneity) can be smooth.

The fluctuations in the cosmic metric (related to the gravitational potential in the Newtonian concept) are indeed quite small in the current cosmological paradigm. These are around $10^{-4}$ level or even less in nearly all cosmic scales except near compact objects like neutron stars and black holes, which are astrophysical subjects rather than cosmology. The strength of metric fluctuations is characterized by dimensionless combinations $GM/(\ell c^2)$ and $v^2/c^2$ where $M$, $\ell$, and $v$ are characteristic scales of the mass, length, and velocity involved in the system. Thus, in the spacetime geometry, the deviations from Minkowski or Robertson-Walker metric are extremely small ($10^{-4}$ level or less) in all cosmological scales. This is the case {\it independently} of high inhomogeneity or nonlinearity in the matter distribution, which is related to the curvature, the second derivative of the metric fluctuations.

We can show the difference between geometry and matter as follows. The Poisson's equation (some limit of the energy constraint equation in Einstein's gravity) in expanding space can be written as $a^{-2} \Delta \delta \Phi = 4 \pi G \delta \varrho \sim H^2 \delta$ with $a$ the cosmic scale factor, $H \equiv \dot a /a$ the Hubble-Lema\^itre parameter, $\delta \equiv \delta \varrho/\varrho$ a dimensionless density parameter and $\delta \Phi$ a perturbed gravitational potential (related to the perturbed metric as $\delta \Phi/c^2$); the intrinsic curvature is of order $a^{-2} \Delta \Phi/c^2$. With $\Delta/a^2 \sim \ell^{-2}$ ($\ell$ a length scale considered), the dimensionless potential (metric) fluctuation becomes $G \delta M/(\ell c^2) \sim \delta \Phi /c^2 \sim (\ell^2 H^2/c^2) \delta \sim (\ell/\ell_H)^2 \delta$, with $\ell_H \equiv c/H$ the horizon scale. Thus, far inside the horizon ($\ell \ll \ell_H \sim 3~h^{-1}\textrm{Gpc}$ with the present value of $H_0 \equiv 100 h$ $\textrm{km}\textrm{s}^{-1}\textrm{Mpc}^{-1}$), the density fluctuation can be significant while the metric fluctuation remains quite small. This is the case we encounter in the CMASS observations via light in this work.

Thus, we understand that while the CP is valid (in the metric!), the density fluctuation is free from the same constraint far inside the horizon. The remaining task is to check whether the observed fluctuations in the matter (or light) are consistent with the theory based on the CP.

The gravitational instability causes the matter inhomogeneity to grow in time even in an expanding background \citep{lifshitz1946}. Thus, the early universe is closer to homogeneity but should still have some inhomogeneity as the seeds for later growth into the large-scale structure. The success of modern cosmology is built on the assumption that the metric fluctuations are small enough. Thus, the CP is imposed in the spacetime metric. In this way, a linear stability analysis is sufficient to handle the formation and evolution processes of the large-scale structure in the early universe and even currently in the large-scale. If the deviations are small (say linear, meaning we can safely ignore the nonlinear terms in the process), the fluctuations cancel out by average over space so that the CP applies as the background. This is the CP adopted in modern cosmology: a spatially homogeneous and isotropic {\it background} plus small enough (preferably linear) perturbations in geometry such that the background remains meaningful. Even in nonlinear perturbation theory, the background is assumed to be intact \citep{hwang2013}. In this way, the success of (linear) perturbation theory in the {\it metric}, instead of actually taking the spatial average over the metric, can be regarded as the validity of the CP in the background.

If the deviations are not small (i.e., the nonlinear terms in the metric cannot be ignored), however, the fluctuations may not disappear by the average, and the background universe where the CP is supposed to apply may lose its meaning. In such a case Einstein's CP may need modifications. Historically there are some manageable (mainly only to the background model, though) examples of inhomogeneous or anisotropic alternative CPs \citep{krasinski1997,ellis2012}. The extreme isotropy of the CMB should severely constrain the homogeneous but anisotropic cosmological models. Another example is the inhomogeneous but spherically symmetric Lema\^itre(-Tolman) model \citep{lemaitre1933} adopting Lema\^itre's CP. As the observations become more precise, finding situations where metric nonlinearities (i.e., deviations from the CP) are important in the radial direction in the cosmic scale will be an interesting discovery. No such a possibility is currently identified {\it within} modern cosmology \citep{hwang2006,jeong2011}, meaning the CP is {\it consistent} with observations.

The nonlinear evolution process of the large-scale structure (of baryon and dark matter) occurs in the matter-dominated era (with the cosmological constant) far inside the horizon. The numerical simulation based on Newtonian gravity is assumed to be sufficient to handle this stage \citep{springel2005,horizonrun2009,teyssier2009,prada2012,angulo2012,heitmann2015,horizonrun2015,habib2016,potter2017,vogelsberger2020}. The Newtonian gravity already {\it implies} that the metric fluctuations deviating from the CP are negligibly small. This is the case independently of whether the process is nonlinear or not and how large the inhomogeneity is. The Newtonian theory is, in fact, a $c$-goes-to-infinity limit of Einstein's gravity. Thus $GM/(\ell c^2)$ and $v^2/c^2$ are naturally {\it negligibly} small. The first-order consideration of these terms coming from Einstein's gravity is known as the post-Newtonian corrections, and currently, these are estimated to be negligibly small in cosmology \citep{shibata1995,hwang2008}. The Poisson equation shows that, as the scale becomes smaller than the horizon scale, the amplitude of inhomogeneity in matter distributions $\delta$ is detached from the potential (metric) fluctuation $\delta \Phi/c^2$.

In this work, we will show that the CMASS distribution of galaxies is {\it consistent} with results from the numerical simulation and the theoretical estimation of matter power spectrum; we use mock galaxy catalogs derived from the Horizon Run 3 simulation \citep{horizonrun2011} and the log-normal distributions of matter density \citep{lognormal2017}. Thus, the CP cannot be found even in objects identified from the cosmological theory based on the CP. We note that the numerical simulation handling the nonlinear process in the Newtonian context is directly built on the CP. The simulation is performed in a comoving coordinate where the background expansion is {\it subtracted} using equations from Einstein's gravity (or modified ones if needed) combined with the CP. The fluctuations in the dimensionless gravitational potential from the Horizon Run 4 simulation \citep{horizonrun2015} show dispersions of $\delta \Phi/c^2$ smaller than $10^{-4}$ in all scales we are considering \citep{horizonrun2021}, thus extremely small indeed.

%%%%%%%%%%%%%%%%%%%%%%%%%%%%%%%%%%%%%%%%%%%%%%%%%%%%%%%%%%%%%%%
%
%
%
%%%%%%%%%%%%%%%%%%%%%%%%%%%%%%%%%%%%%%%%%%%%%%%%%%%%%%%%%%%%%%%
\subsection{The ontological status of the CP}

Therefore, as long as we accept the Robertson-Walker metric as the background, the HS covers {\it all} cosmological scales concerning the metric fluctuations. The CP applies as long as we can ignore nonlinearity in the metric fluctuations, and the metric inhomogeneity in current cosmology is estimated to be $10^{-4}$ level or less in all cosmologically relevant scales. As the CP in modern cosmology concerns the metric only, searching for the HS in the matter or light is {\it not relevant} to the CP. In Section \ref{sec:test}, using the CMASS sample, we will show that the HS cannot be found up to the colossal $300~h^{-1}\textrm{Mpc}$ in radius; we already have an example in the SDSS Sloan Great Wall showing a dramatic structure! But the distribution is consistent with theoretical results derived from the CP, and even bigger structures are expected in the simulation, see \citet{cpark2012}.

Our study shows that the CP can be found in neither matter nor light; up to the scale at which the current observation is available. Thus, the CP is not embedded in Nature. The CP, however, is well valid in the metric. The metric fluctuations are sufficiently small in the observed cosmic scale, and this is where the CP is needed in theoretical cosmology; the CP (in the metric) is perfectly reasonable within modern cosmology (which already assumes the CP as the background, though). Here is a subtle difference between the theory and observation. The metric (or potential) is a theoretical concept not directly observable. This reveals the ontological status of the CP, a prescient theoretical postulation by Einstein, making all the current success of modern cosmology possible.

%%%%%%%%%%%%%%%%%%%%%%%%%%%%%%%%%%%%%%%%%%%%%%%%%%%%%%%%%%%%%%%
%
%
%
%%%%%%%%%%%%%%%%%%%%%%%%%%%%%%%%%%%%%%%%%%%%%%%%%%%%%%%%%%%%%%%
\section{Test}\label{sec:test}

In this study, as a fiducial model we take a $\Lambda$CDM model with the current matter and dark energy density parameters $\Omega_{m} = 0.27$ and $\Omega_{\Lambda} = 0.73$, respectively. Here the dark energy is based on Einstein's cosmological constant $\Lambda$. 

\begin{figure}[!]
\includegraphics[width=\columnwidth,trim=0 0 0 0]{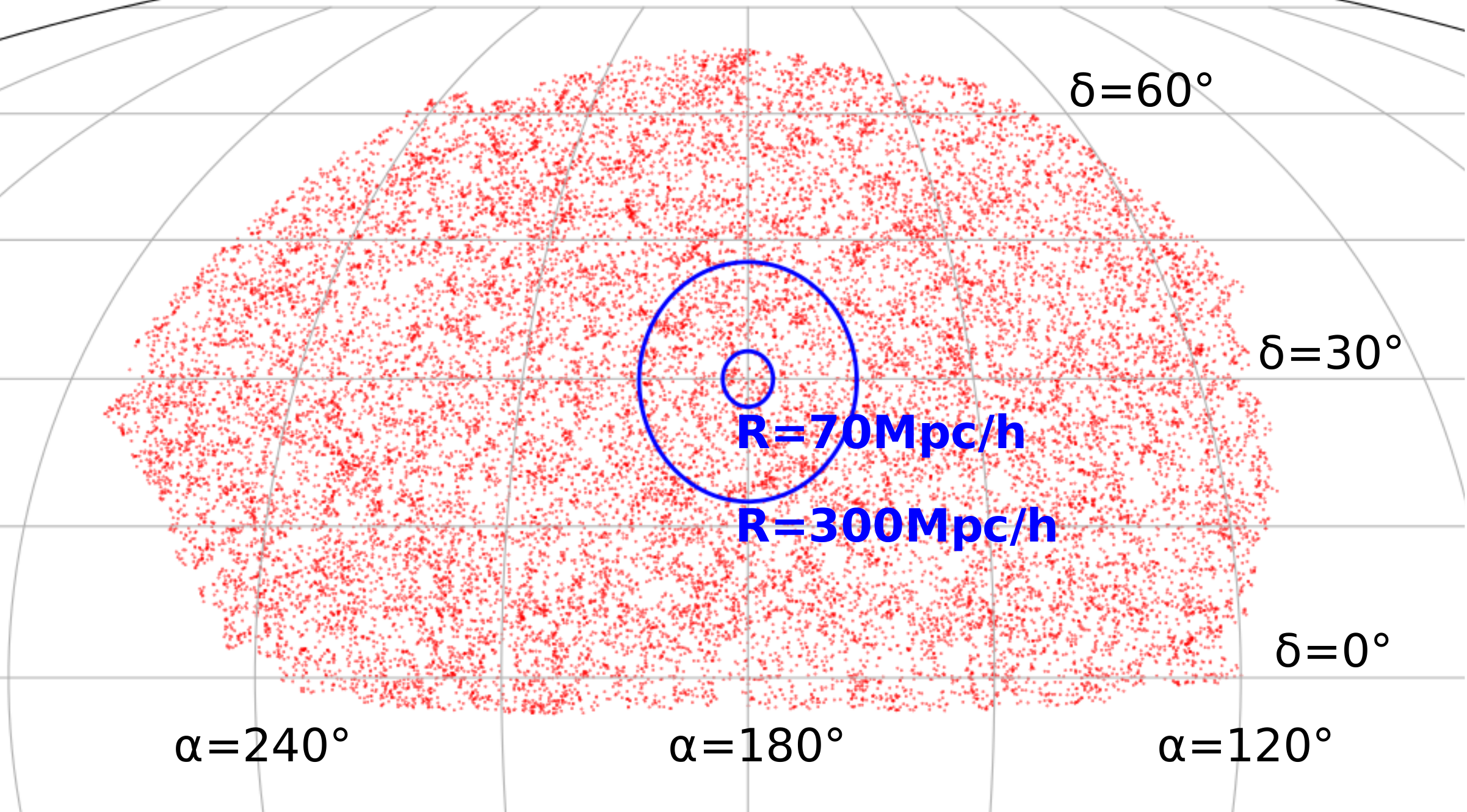}
\caption{Angular distribution of CMASS galaxies of the Northern Galactic Cap within a slice of $100~h^{-1}\textrm{Mpc}$ thickness centered at the redshift $z_{\rm{cen}} = 0.52$, shown in Mollweide equal-area projection in the equatorial coordinate system ($\alpha$ and $\delta$ denote right ascension and declination, respectively). In order to increase visibility so that the distribution of galaxies is clearly visible, only 14\% of the total data points were randomly selected and displayed. For a comparison of spatial scales, two circles with radii of $70$ and $300h^{-1}\textrm{Mpc}$ have been added.}
\label{fig:CMASS-distribution}
\end{figure}

\subsection{Data}\label{sec:data}

\subsubsection{BOSS CMASS sample}

We used the CMASS sample of massive galaxies from the 12th data release of SDSS-III Baryonic Oscillation Spectroscopic Survey (BOSS). The target galaxies in the CMASS sample are selected so that the stellar mass of the systems is nearly constant at a redshift range of $0.43 < z< 0.7$ \citep{eisenstein2011, anderson2012, cbias2}. The CMASS sample is roughly a volume-limited sample up to the redshift $z=0.5$, but after that it is close to a magnitude-limited sample.
The resulting sample is composed of luminous galaxies with approximately uniform comoving space density of $3\times 10^{-4} h^3 \textrm{Mpc}^{-3}$. Although the CMASS sample does not represent all the galaxies due to various color-magnitude cuts applied to target galaxies with a certain limit of stellar mass, it is more suitable for testing homogeneity of matter distribution compared to the former LRG sample \citep{eisenstein2001, abazajian2009, kazin2010} in the sense that it provides a less biased sample of massive galaxies by allowing galaxies with lower luminosity and wider color range \citep{dawson2012}. It provides a sufficient number of galaxies with the photometric and spectroscopic information, a total of $931,517$ galaxies in the Northern Galactic Cap (NGC) and the Southern Galactic Cap (SGC) that covers $9,376~\textrm{deg}^{2}$, about one fourth of the whole sky \citep{alam2015}.

\begin{figure}[!]
\includegraphics[width=\columnwidth,trim=0 0 0 0]{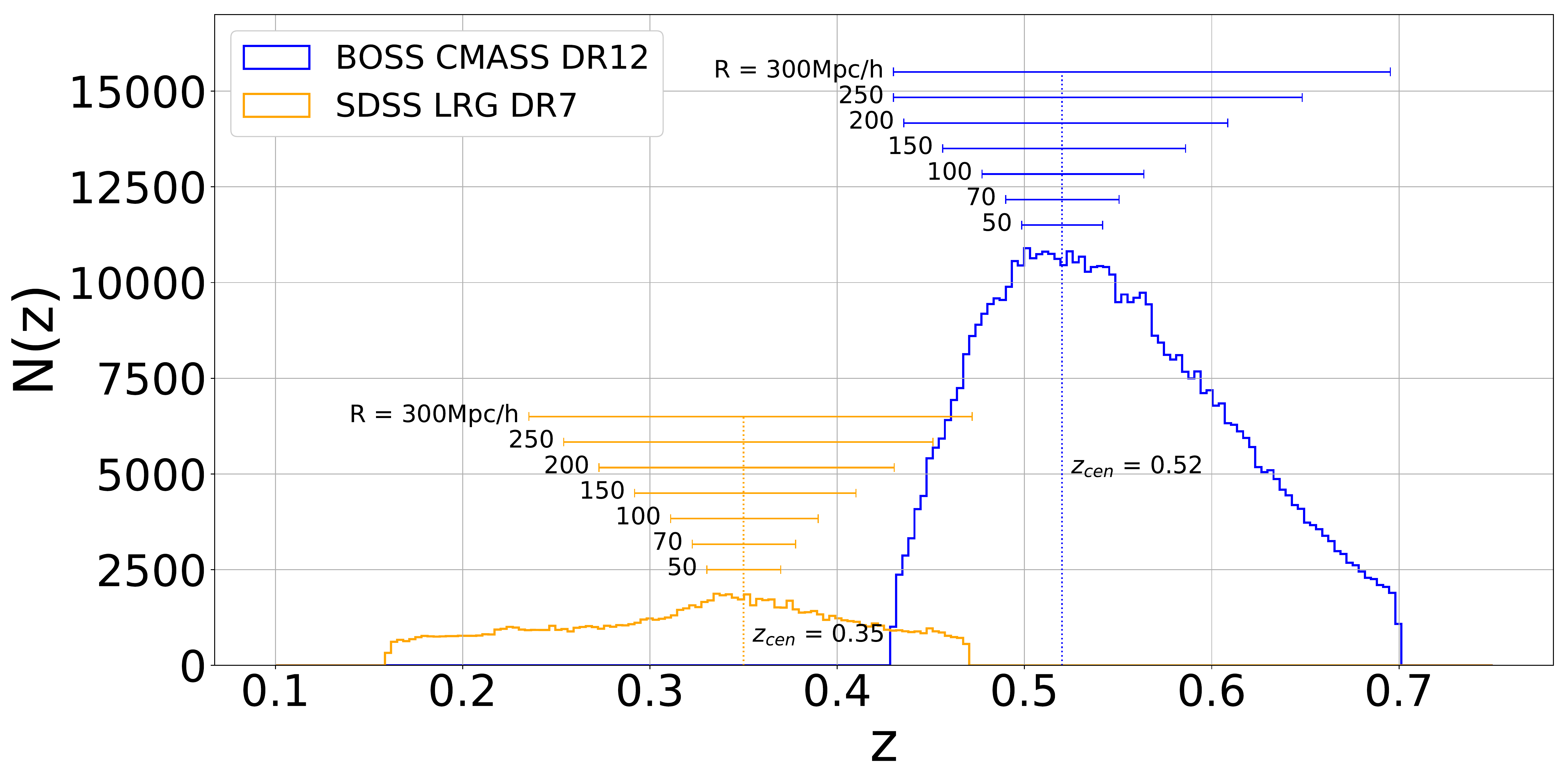}
\caption{Number distributions of the LRG and CMASS galaxies over redshift. The width of redshift in the histograms is $\Delta z = 3.25\times 10^{-3}$.  Also shown are the redshift ranges enclosed by spheres with a series of radii from $50$ to $300~h^{-1}\textrm{Mpc}$, centered at $z_{\rm{cen}} = 0.35$ for LRG and $z_{\rm{cen}} = 0.52$ for CMASS (denoted as vertical dotted lines). For radii of $250$ and $300~h^{-1}\textrm{Mpc}$ in the case of the CMASS sample, the redshift ranges have been shifted to the higher redshift to include more galaxies within the range specified.}
 \label{fig:CMASS-radial}
\end{figure}

In this paper, we use only $568,776$ galaxies contained in the NGC (Fig.\ \ref{fig:CMASS-distribution}). This is five times larger than the number of LRG galaxies used in the previous study \citep{cgpark2017}, enabling more reliable statistical calculations. Figure \ref{fig:CMASS-radial} shows the radial distribution of the CMASS galaxies used in this work, together with that of the LRG galaxies. The redshift ranges enclosed by spheres that are located at the central redshift of both surveys are also shown for a series of radii from $50$ to $300~h^{-1}\textrm{Mpc}$.
Our fiducial $\Lambda\textrm{CDM}$ model has been used to transform the redshift to comoving distance.
We will use the spheres for the CMASS sample to define the truncated cones with characteristic radius scale and count CMASS galaxies within them for testing homogeneity. The CMASS radial number distribution will be used to make random and mock catalogs so that the data points can be located similarly to those in the CMASS sample.

\begin{figure}
\includegraphics[width=\columnwidth,trim=0 0 0 0]{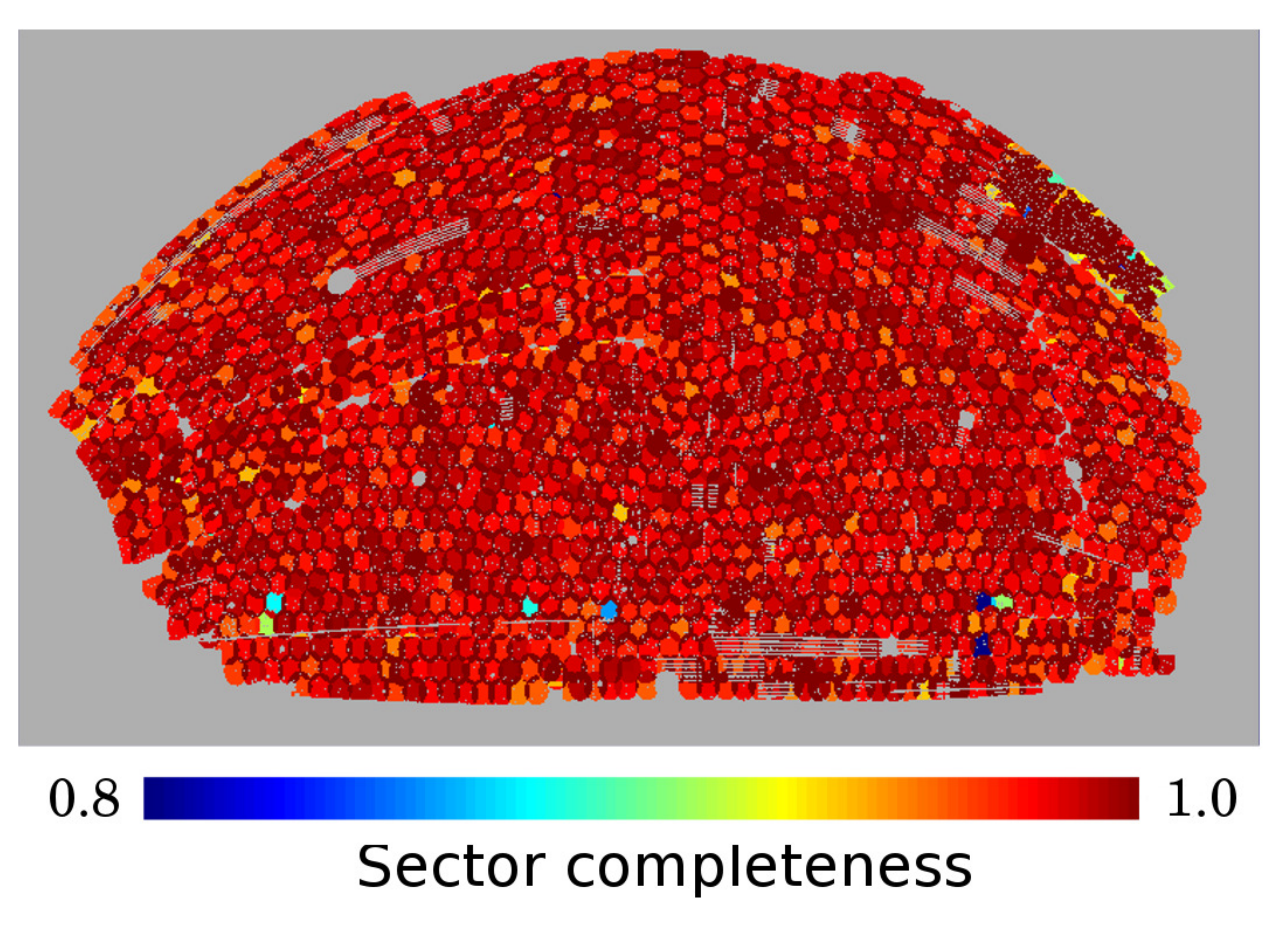}
\caption{The sector completeness map of the BOSS DR12 CMASS sample. The grey color indicates the outer region of the survey or no data.}
 \label{fig:CMASS-ASF}
\end{figure}

Due to the mask of bright stars, centerposts of telescope, observation failure, and a number of various reasons, the galaxies within the BOSS survey region were not completely surveyed, and thus the missing and incomplete regions exist. We reconstruct the completeness of the survey by combining the BOSS survey boundaries and a collection of mask information provided by the BOSS website  \footnote{The SDSS Science Archive Server, https://data.sdss.org/sas/dr12/boss/lss/geometry/}. The MANGLE software is used to interpret and combine the mask files with polygon format \citep{mangle1, mangle2}. The final output of the MANGLE for the completeness of the CMASS sample is transformed into the HEALPix format where the individual pixels cover equal area on the sky \citep{healpix}. The CMASS sector completeness within the BOSS survey region is shown in Fig.\ \ref{fig:CMASS-ASF}. Hereafter, we call this completeness map as the angular selection function (ASF) of the CMASS sample.

In our analysis, we take into account the fact that different weights are assigned to galaxies that were placed in various observation environments. For example, a galaxy is given more weight if the nearest neighbor of this galaxy is located within the diameter of the spectroscopic fiber and the redshift of the neighboring galaxy was not obtained due to fiber collision. We correct for the effect of the fiber collision ($w_{\rm{cp}}$) and the redshift failure ($w_{\rm{zf}}$). Additionally, we also apply weights to account for the systematic relation between the number density of observed galaxies and stellar density ($w_{\rm{star}}$) and seeing ($w_{\rm{see}}$). Finally, each galaxy is assigned the combined weight defined as $w_{\rm{tot}} = (w_{\rm{cp}}+w_{\rm{zf}}-1) w_{\rm{star}} w_{\rm{see}}$. By combining the survey completeness ($w_{\rm{sc}}$) at the angular position of a galaxy (Fig.\ \ref{fig:CMASS-ASF}) and the weights assigned to it, we obtain the contribution of each galaxy in the galaxy count as $w_{\rm{tot}}/w_{\rm{sc}}$.

\subsubsection{Random catalogs}

Before testing the homogeneity of matter distribution using the CMASS galaxies, first we need to establish a clear {\it criterion} for determining homogeneity. Although homogeneity implies that the physical state is the same at all locations, in principle it is impossible for a finite number of points within a space to be homogeneously distributed. This is because fictitious clustering and density inhomogeneity are always visible when a finite number of data points are generated even with uniform probability. Therefore, here we establish the criterion for homogeneity from a statistical point of view. The simplest way to do this is to use the distribution of random points following the Poisson distribution as the criterion for homogeneity.
The finite number of points randomly distributed by the Poisson distribution shows fluctuations in number density from place to place. On a small scale, the actual distribution of galaxies tends to show a larger fluctuation due to clustering of galaxies than expected in the Poisson distribution. Since these fluctuations are expected to decrease with increasing scale, we can determine the HS from which the distribution of galaxies begins to match the random distribution within a statistically acceptable range.

In this work, we have made one thousand random catalogs by considering the observational characteristics of the CMASS sample, each containing data points equal to the number of CMASS galaxies. The random data points are generated within the survey region using the radial number distribution and sector completeness as the probability density functions.
However, the random data points were assigned uniform weight ($w_{\rm{tot}}=1$), unlike the galaxies with various weights. Thus, the contribution of each point to the count is $1/w_{\rm{sc}}$ for the random catalog. We analyzed these random catalogs in the same way as the CMASS sample is analyzed.

\subsubsection{HR3 mock catalogs}

To test the homogeneity of the CMASS galaxy distribution, in addition to the random distribution, we also analyzed the mock galaxy distribution expected in the standard cosmological model, and compared the results with that of the actual galaxy distribution. Here we have produced the CMASS mock catalogs from the two model-derived data sets.

\begin{figure}
\includegraphics[width=\columnwidth,trim=50 20 110 50]{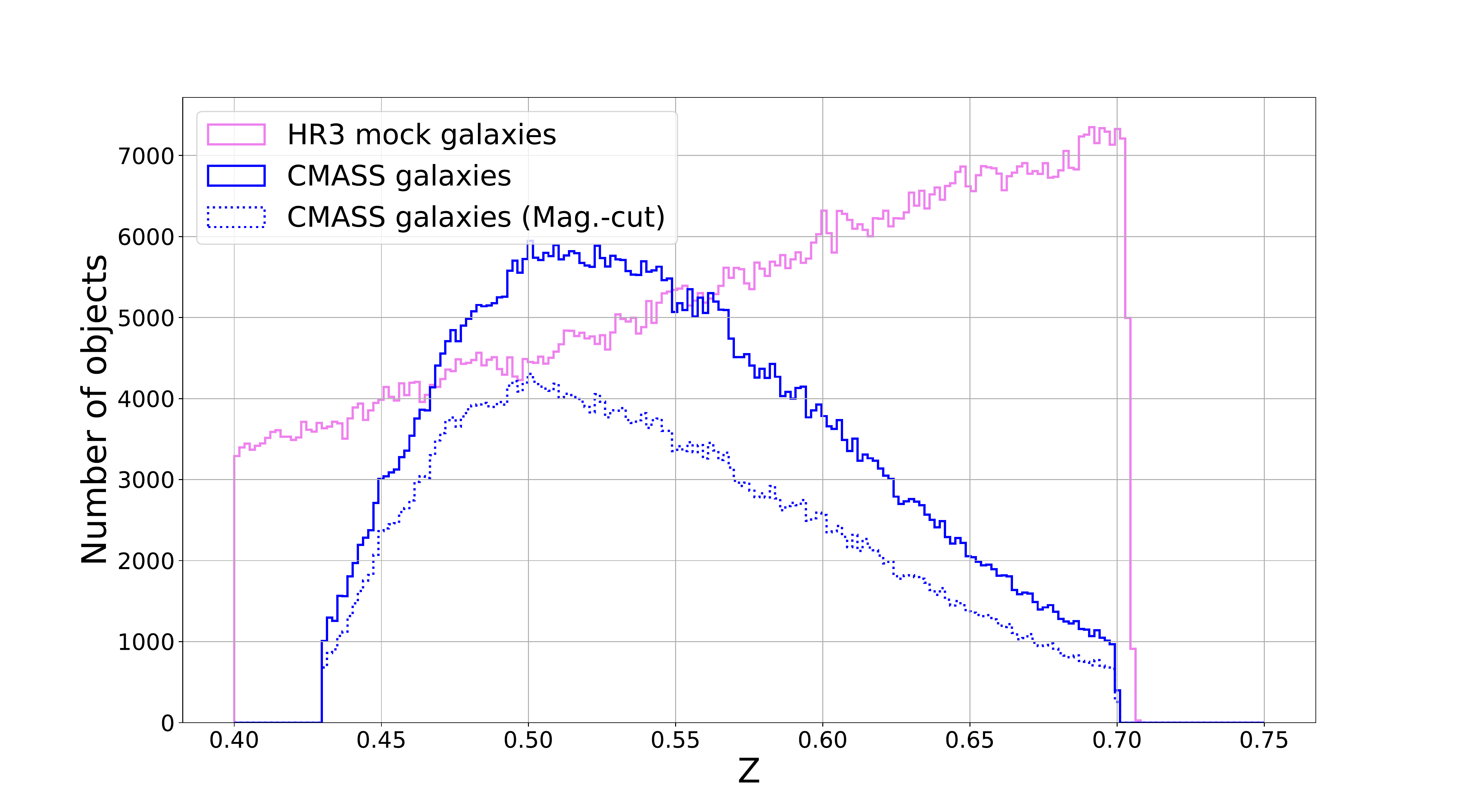}
\caption{Radial number distributions of Horizon Run 3 (HR3) mock galaxies ($N=918,135$) from the first of 27 lightcones (only ASF is applied), the CMASS galaxies ($N=568,776$), and the CMASS galaxies brighter than the absolute magnitude of $-20.2$ in $r$-band ($N=395,011$).}
\label{fig:newz}
\end{figure}

The first one we use is the 27 all-sky mock BOSS survey data provided by Korea Institute of Advanced Study\footnote{Available at http://sdss.kias.re.kr/astro/Horizon-Runs/Horizon-Run23.php}. The data were derived from the Horizon Run 3 (HR3) simulation, one of the largest $N$-body simulation in volume to date with $7,210^{3}$ dark matter particles within the volume of $(10.815~h^{-1} {\rm{Gpc}})^{3}$ (\citealt{horizonrun2011}). The halos were picked up as mock galaxies at redshift range $0.43 < z < 0.7$ considering the redshift-dependent mass limit determined from the expected galaxy number density of the BOSS survey (see $\S4.2$ of  \citealt{horizonrun2011}).
Figure \ref{fig:newz} compares the radial number distributions of HR3 mock galaxies (extracted from one of the mock BOSS survey data) and the CMASS galaxies. For HR3 data, the mock galaxies corresponding to the percentage of CMASS sector completeness have been randomly selected within the survey region. Noting that the radial number distribution of the selected HR3 mock galaxies is less than that of the CMASS sample at redshift range of $0.47 < z < 0.55$, we have excluded the CMASS galaxies that are fainter than absolute magnitude of $-20.2$ in $r$-band and made the galaxy number density around $z=0.5$ smaller than the HR3 data so that mock catalogs for the CMASS sample can be safely obtained.
Within the CMASS survey region, considering the CMASS radial number distribution and angular selection function, we extracted massive halos with the same number of the magnitude-limited CMASS galaxies in decreasing order of halo mass.
For each mock SDSS-III survey data that covers the whole sky, it is possible to extract the four independent mock catalogs for the CMASS NGC region. Thus, we have made a total of $108$ independent mock catalogs. The limited number of mock catalogs makes it possible to judge statistical significance only up to $2\sigma$.
Each HR3 mock catalog contains $395,011$ data points. We also have made separate sets of random catalogs that match the new radial number distribution of the magnitude-limited CMASS sample (see Fig.\ \ref{fig:newz}).

Our method of generating mock catalogs has a limitation in that it cannot specifically mimic the process in which CMASS galaxies are selected by various color-cuts. However, our method is valid in that it mimicks massive galaxies of the CMASS sample by preferentially choosing heavier halos, taking into account the known relation that a more massive halo contains a brighter galaxy \citep{kim-etal-2008}.

\subsubsection{Log-normal mock catalogs}

The second data is the mock galaxies extracted from the log-normal simulations, which is an inexpensive way of generating realizations of nonlinear density fluctuations. We use a publicly available code for generating log-normal realizations of galaxies in redshift-space\footnote{https://bitbucket.org/komatsu5147/lognormal\_galaxies/src/master/} \citep{lognormal2017}. The code generates log-normal matter density field from matter power spectrum of the $\Lambda$CDM model, then draws Poisson-distributed mock galaxies with a number expected by the log-normal density field at a given position in space. It has been confirmed whether the generated mock galaxies give a two-point statistics similar to that of a standard $\Lambda$CDM model \citep{lognormal2017}.

Using the flat $\Lambda$CDM model favored by the Planck 2018 TT,TE,EE+lowE+lensing data (\citealt{planck2020}), we obtain the log-normal realizations of galaxies at redshift $z_{\textrm{cen}}=0.52$. Since galaxies in the log-normal realizations do not have mass information, there is a disadvantage that galaxies cannot be selected in order of mass as in the HR3 mock catalog. To obtain the statistics similar to CMASS galaxies, we assume the bias of galaxies relative to dark matter to be $b=2.2$, which is a value adopted by BOSS collaboration for generating mock catalogs \citep{alam2017}. This value does not statistically deviate from the measurement of other studies \citep{guo2013, cbias2, lbias3}. Finally, we generated one thousand mock catalogs by randomly picking up mock galaxies in the log-normal realizations considering the radial number distribution and sector completeness of the CMASS sample so that the generated log-normal galaxies follow the magnitude-limited CMASS galaxies. We also produced 1,000 log-normal catalogs even assuming a lower bias of $b=1.8$ as suggested in \citet{cbias1}. It is shown that the log-normal catalogs with $b=1.8$ do not provide statistics similar to the CMASS data (see Fig.\ \ref{fig:CMASS-stat} below). 

As in the random catalogs, we assigned the uniform weights ($w_{\rm{tot}} = 1$) for all points in both the HR3 mock catalogs and the log-normal mock catalogs.

%%%%%%%%%%%%%%%%%%%%%%%%%%%%%%%%%%%%%%%%%%%%%%%%%%%%%%%%%%%%%%%
%
%
%
%%%%%%%%%%%%%%%%%%%%%%%%%%%%%%%%%%%%%%%%%%%%%%%%%%%%%%%%%%%%%%%
\subsection{Method: Counting galaxies with truncated cones}\label{sec:method}

\begin{figure}
\includegraphics[width=\columnwidth,trim=0 0 0 0]{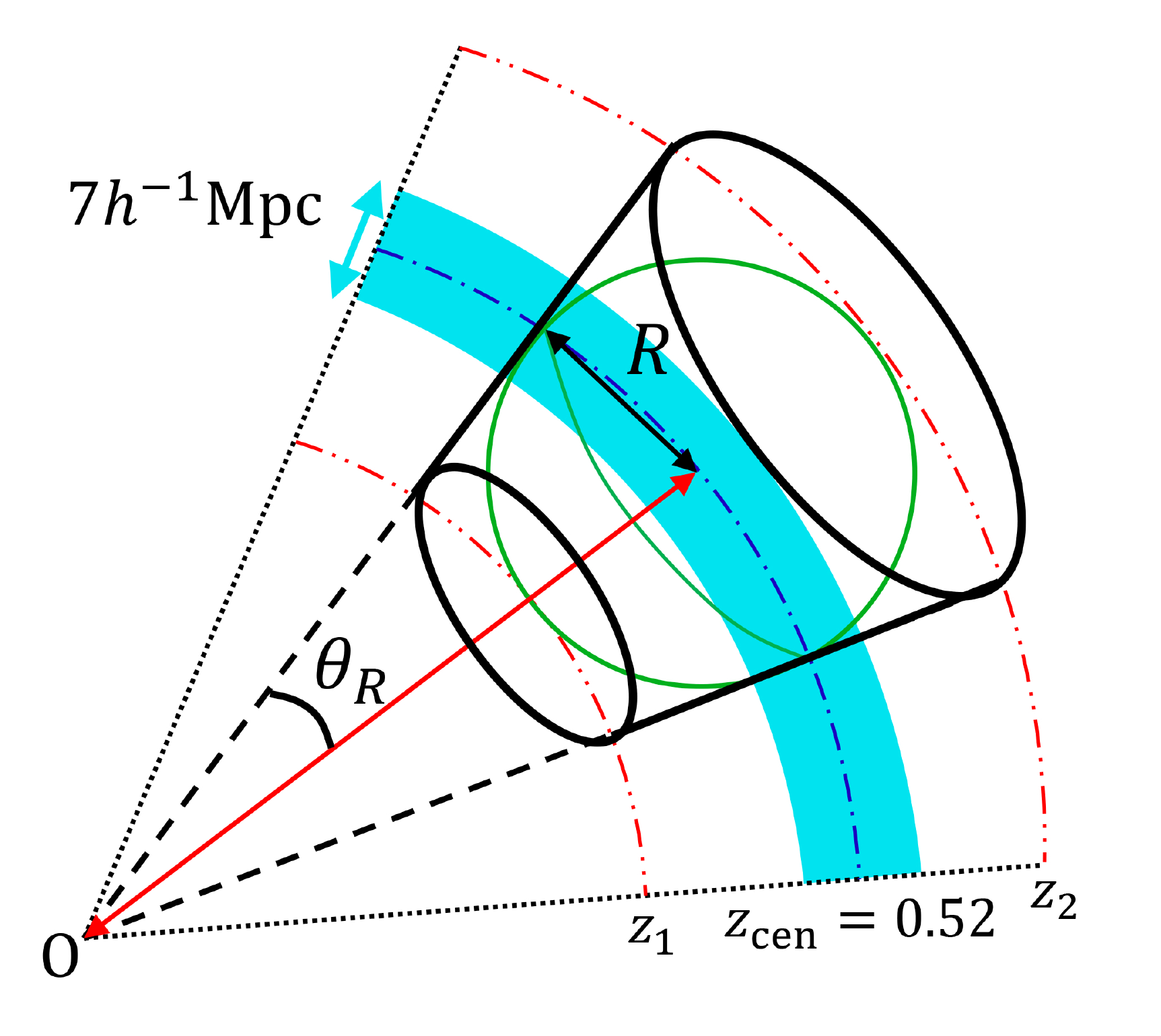}
\caption{A truncated cone circumscribed about a sphere of radius $R$. The apparent angular radius of the truncated cone (or the sphere) as seen from the observer at O is $\theta_R$. Redshifts $z_1$ and $z_2$, the lower and upper limits of the truncated cone, respectively, vary as the radius of the sphere at the central redshift $z_{\rm{cen}}$ varies. The thin slice at $z_{\rm{cen}}$ has thickness of $\Delta r = 7 h^{-1}\textrm{Mpc}$. We count galaxies inside the truncated cone that is centered on every galaxy within the slice.}
\label{fig:trccone}
\end{figure}

To quantitatively test whether the three-dimensional spatial distribution of galaxies is homogeneous, we have applied the galaxy counting method to the CMASS sample, random catalogs, and two mock catalogs.
\citet{hogg2005} and \citet{netlis2017} used the count-in-sphere method to judge whether the distribution of galaxies is homogeneous by comparing the number of galaxies within a sphere of specific radius with that of the random distributions.
On the other hand, \citet{cgpark2017} tested the homogeneity of the LRG distribution by counting galaxies within a truncated cone instead of the sphere.
Unlike the count-in-sphere method that needs the three-dimensional spatial distribution of galaxies, the count-in-truncated-cone method with fixed redshift center is less dependent on specific cosmological models in that galaxies are counted in the given redshift space, not in the real space. In addition, counting galaxies within a truncated cone enables us to define a criterion for homogeneity without the need to consider different statistical weights given to galaxies that are unevenly distributed along the radial direction. This is because the accumulated statistical weight of the line-of-sight direction is the same no matter which direction we see within the truncated cone.

\begin{figure*}
\includegraphics[width=85mm,trim=0 -90 0 0]{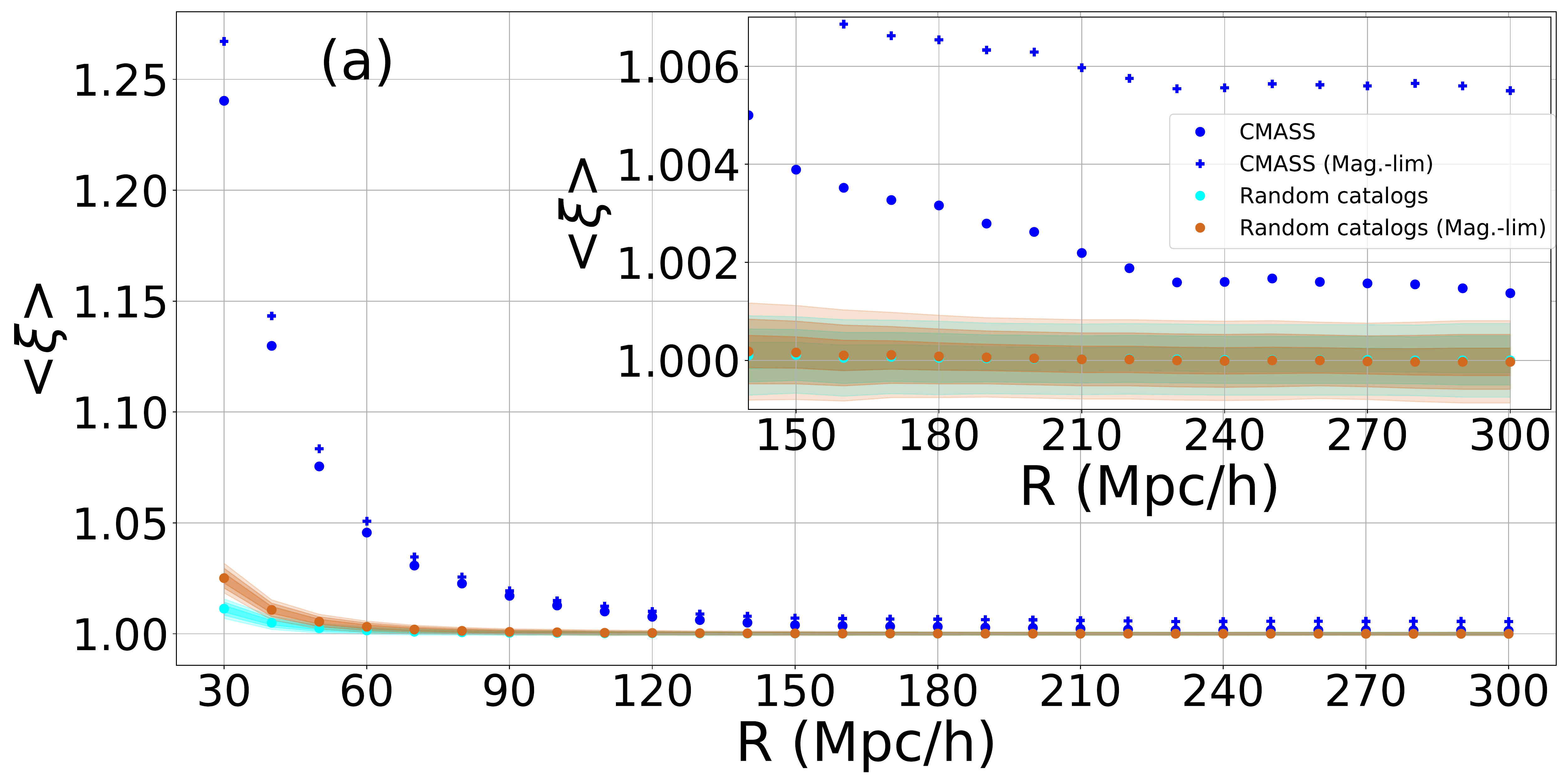}
\includegraphics[width=85mm,trim=0 -90 0 0]{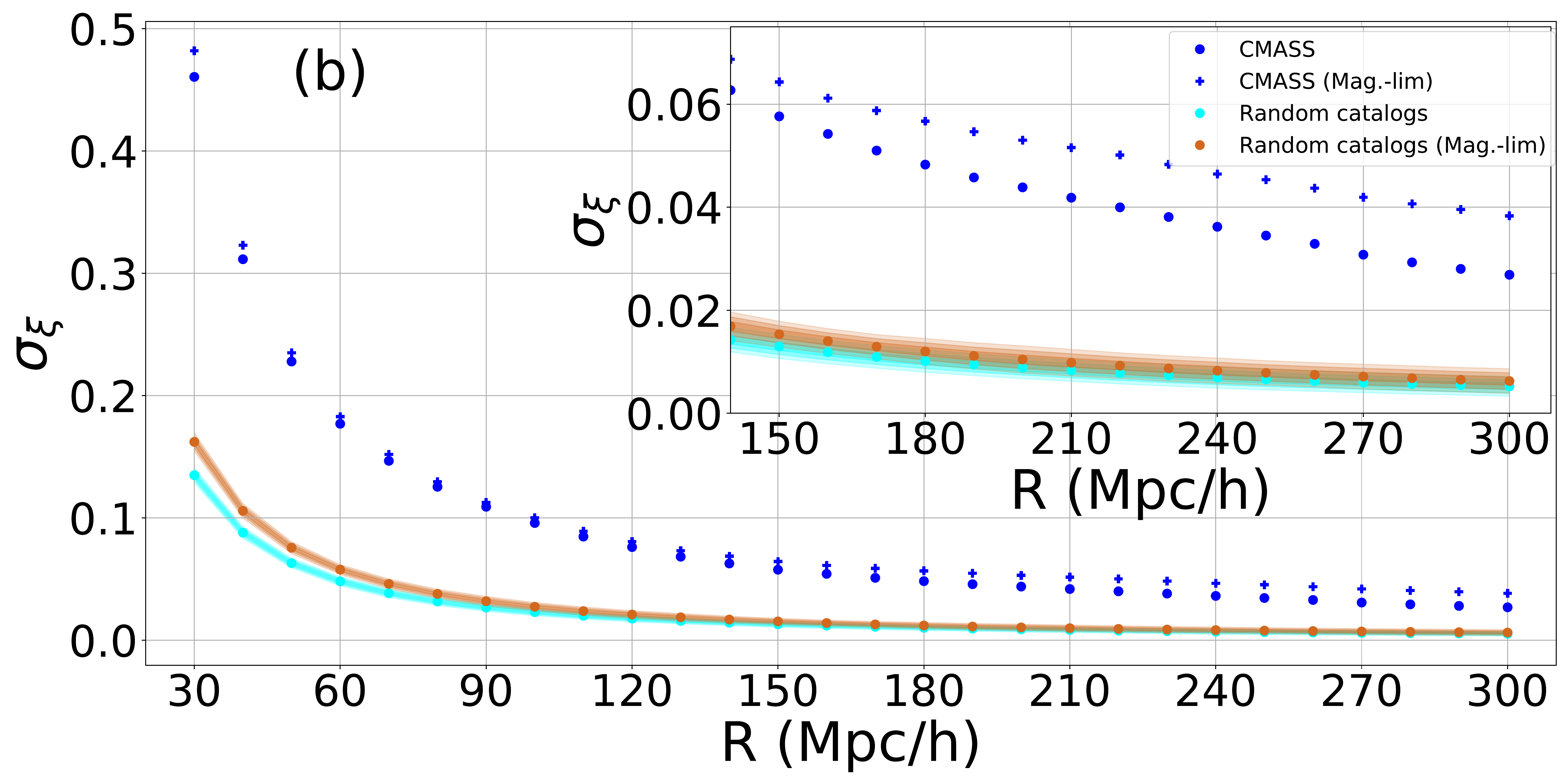}\\
\includegraphics[width=85mm,trim=0 -90 0 80]{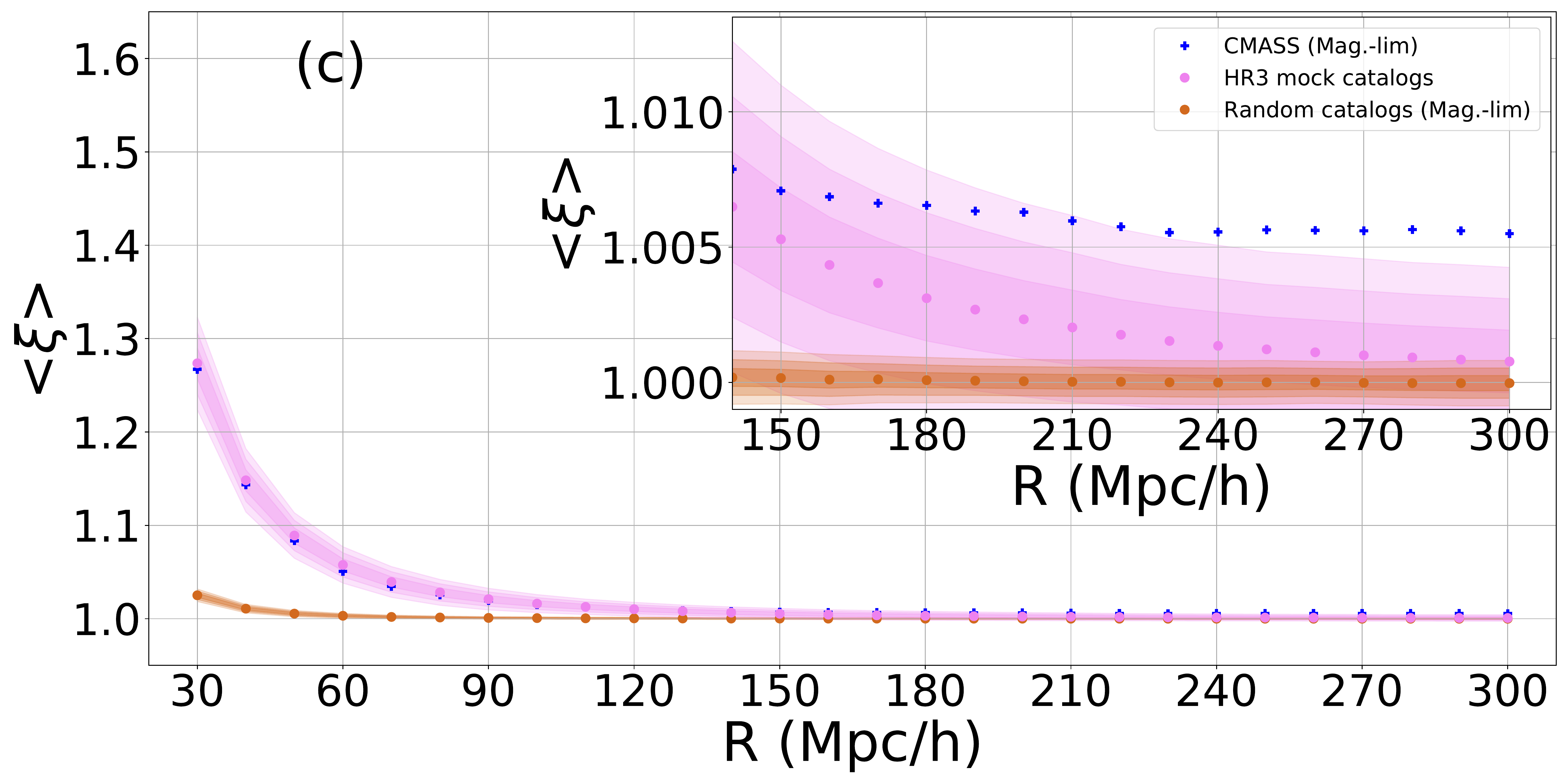}
\includegraphics[width=85mm,trim=0 -90 0 80]{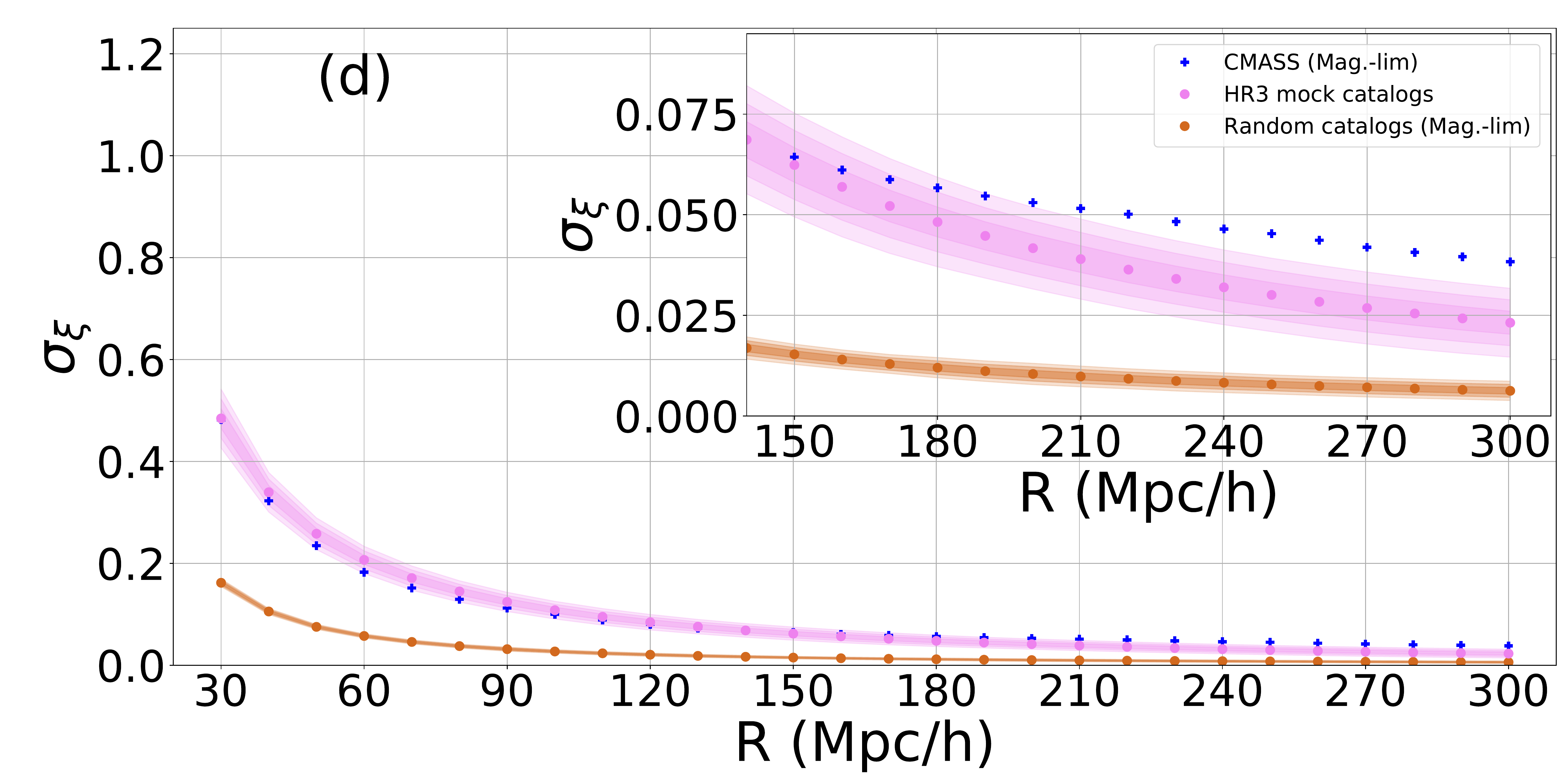}\\
\includegraphics[width=85mm,trim=0 -10 0 80]{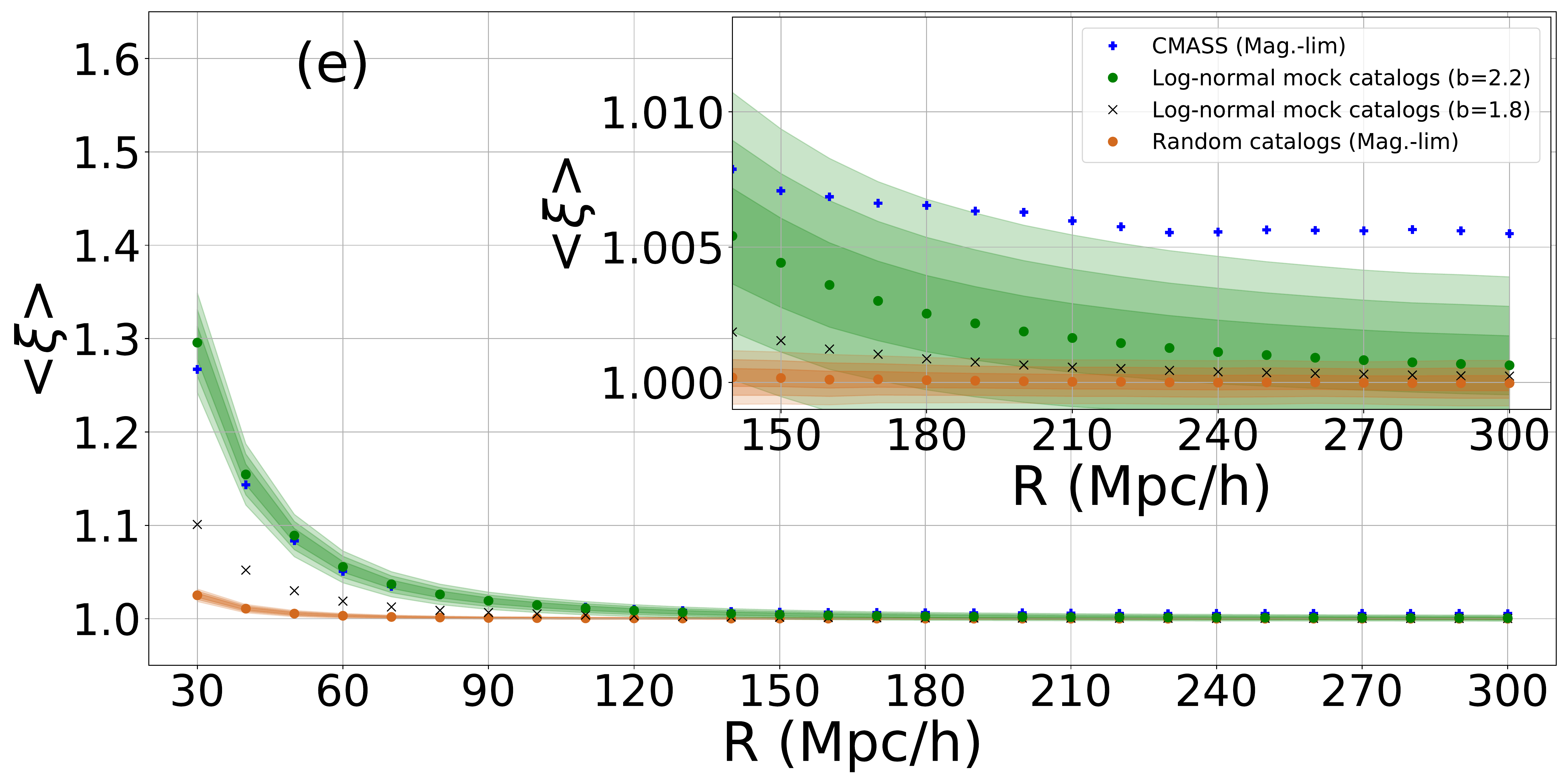}
\includegraphics[width=85mm,trim=0 -10 0 80]{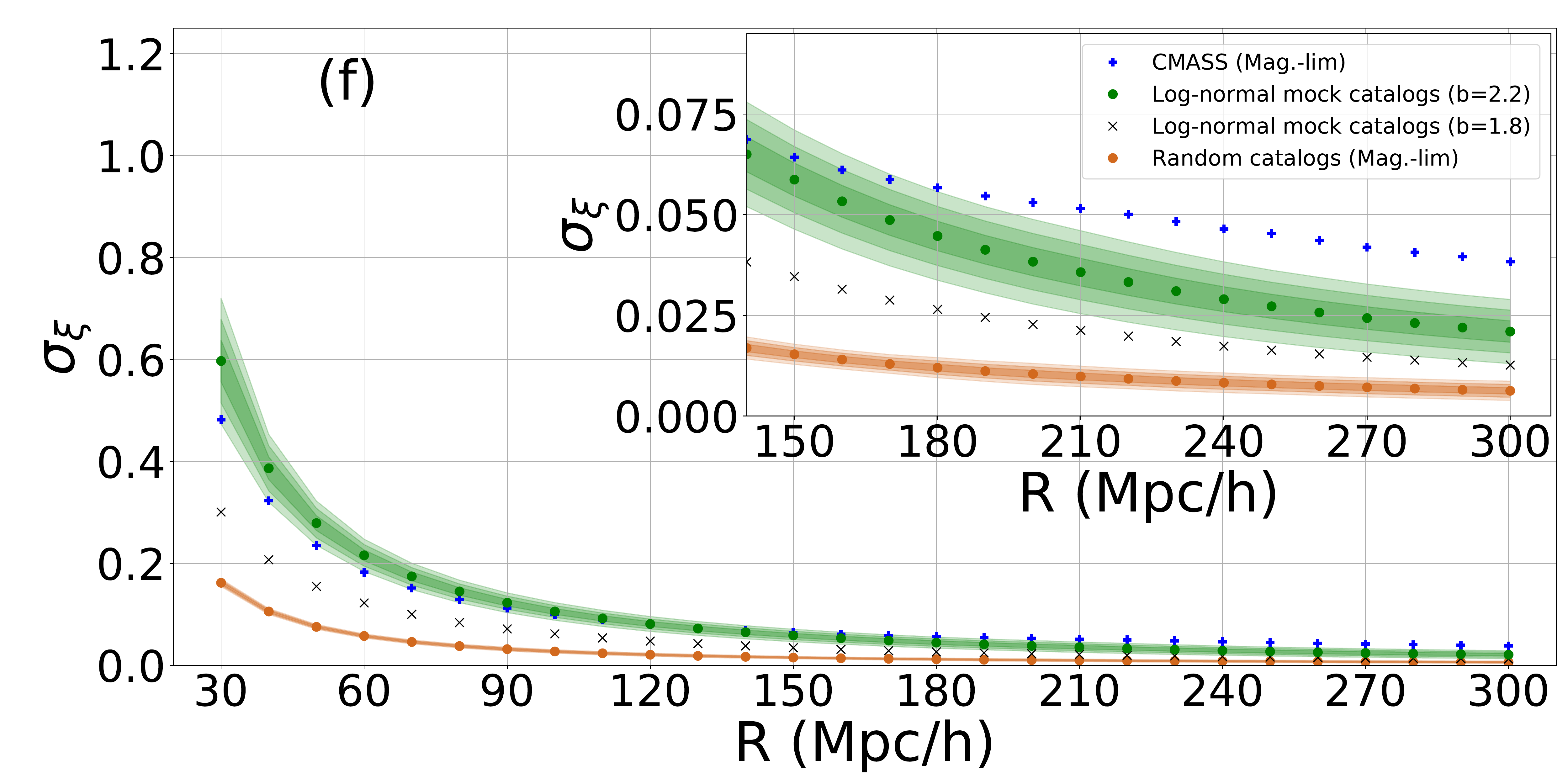}
\caption{The mean ($\left< \xi \right>$) and standard deviation ($\sigma_\xi$) of $\xi$ as a function of radius ($R$). Top panels (a,b) shows the results of unmodified CMASS sample (blue dots) and the corresponding $1,000$ random catalogs (cyan dots), together with those of magnitude-limited CMASS sample (blue pluses) and the corresponding $1,000$ random catalogs (brown dots). The results of 108 Horizon Run 3 (HR3) mock catalogs (pink dots) are shown in the middle panels (c,d), together with those of the magnitude-limited CMASS sample and the corresponding random catalogs. The bottom panels (e,f) presents the results of the log-normal mock catalogs generated with a bias factor $b=2.2$ (green dots). For a comparison, the results for $b=1.8$ are shown as cross-marks ($\times$'s). The shaded regions in each panel, in order from dark to light, represent 1, 2, and 3 sigma confidence regions, which have been estimated from measured means and standard deviations of individual random or mock catalogs. The information of the shaded regions between the specified radius scales are interpolated.}
\label{fig:CMASS-stat}
\end{figure*}

In this work, we use the count-in-truncated-cone method suggested by \citet{cgpark2017}. We consider a $7~h^{-1}\textrm{Mpc}$-thick slice for a central redshift $z_{\rm{cen}}=0.52$, where CMASS galaxies are the most dense. There are $9,770$ CMASS galaxies in this slice. Thickness of the slice was chosen to control the number of galaxies (to be located at the center of the truncated cone) to an appropriate level of about $10,000$. This number is suitable for closely looking at the direction-dependency of galaxy counts within the survey region and for performing calculations within a reasonable time. For each galaxy, we consider a sphere of comoving radius $R$ that is centered at the slice but at the galaxy's angular position, and construct a truncated cone that is circumscribed about the sphere (Fig.\ \ref{fig:trccone}). The apparent angular radius of the truncated cone seen from the observer is given by $\theta_{R}=\sin^{-1}(R/r_{\rm{cen}})$, where $r_{\rm{cen}}$ is the comoving distance to the center of the slice. The redshift $z_1$ ($z_2$) corresponding to the near (far) side of a truncated cone is transformed from the comoving distance $r_{\rm{cen}}-R$ ($r_{\rm{cen}}+R$) in the fiducial $\Lambda$CDM model. For radii from $250$ to $300~h^{-1}\textrm{Mpc}$, however, the $r_{\rm{cen}}-R$ becomes less than the distance to the minimum redshift ($z=0.43$). Thus, in these cases, $z_1$ is set to $z=0.43$ and $z_2$ is set to the higher value so that the interval $z_2-z_1$ can span a range of $2R$ and include more galaxies within the range (see Fig.\ \ref{fig:CMASS-radial}).

Next, we calculate a statistical measure of homogeneity $\xi$, defined as the number density within the truncated cone divided by the number density within the slice covering the whole survey area from redshift $z_1$ to $z_2$,
\begin{equation}
   \xi = \frac{N_{\rm{trc}}/V_{\rm{trc}}}{N_{\rm{slice}}/V_{\rm{slice}}}
\end{equation}
where $N_{\rm{trc}}$ ($N_{\rm{slice}}$) denotes the number of galaxies within the truncated cone (slice) and $V_{\rm{trc}}$ ($V_{\rm{slice}}$) the volume of the truncated cone (slice).
For a homogeneous distribution, the statistic $\xi$ is expected to converge to 1 as the radius increases. However, when the radius is large enough to cover the entire survey region, this statistic has a characteristic of reaching 1 in principle even when the distribution considered is not homogeneous.

Due to the survey boundaries, masks, and holes, a truncated cone is not complete.
The partial volume of the truncated cone is calculated by summing the volume elements corresponding to each HEALpix pixel in the pixelized area occupied by the truncated cone on the sky (see Appendix of \citealt{cgpark2017} for a detailed description).
For varying radius from $R=30~h^{-1}\textrm{Mpc}$ to $300~h^{-1}\textrm{Mpc}$ with steps of $10~h^{-1}\textrm{Mpc}$ (totally 28 steps), we calculate $\xi$ and the volume of the partial truncated cone for every galaxy within the slice with $z=z_{\textrm{cen}}$ and $7~h^{-1}\textrm{Mpc}$ width. The maximum radius of $300~h^{-1}\textrm{Mpc}$ is limited by the depth of the CMASS sample, about $600~h^{-1}\textrm{Mpc}$.

Given individual measurements of $\xi$ for every galaxies within the slice, we calculate the mean ($\left<\xi\right>$), standard deviation ($\sigma_\xi$), skewness ($skew_\xi$), and kurtosis ($kurt_\xi$) of the measured $\xi$'s. During calculating these statistics, we assign each measured $\xi$ a statistical weight by the volume of the partial truncated cone. Thus, the $\xi$'s measured near the survey boundary usually have lower statistical weights than those near the center of the survey region. For each radius scale, these statistics are compared with those estimated from the random and mock catalogs. The significant deviation of the measured value from the expectation of the random or mock distributions may imply the deviation from the homogeneity or the present cosmological model.

%%%%%%%%%%%%%%%%%%%%%%%%%%%%%%%%%%%%%%%%%%%%%%%%%%%%%%%%%%%%%%%
%
%
%
%%%%%%%%%%%%%%%%%%%%%%%%%%%%%%%%%%%%%%%%%%%%%%%%%%%%%%%%%%%%%%%
\begin{figure}
\includegraphics[width=90mm,trim=30 30 20 0]{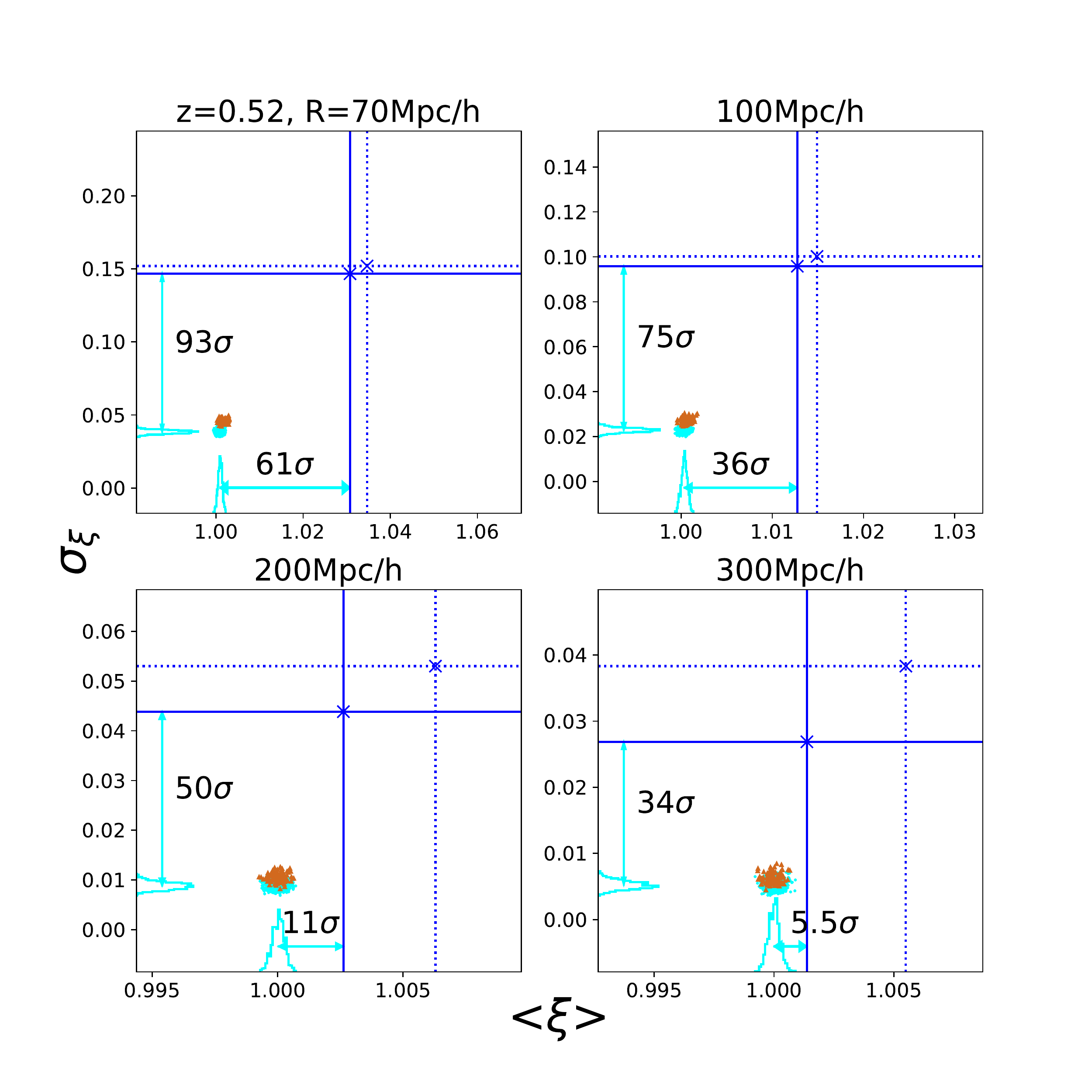}
\caption{The mean and standard deviation diagrams of the unmodified CMASS sample (blue markers and solid lines) and the corresponding $1,000$ random catalogs (cyan dots and histograms) for four different radii at the central redshift $z_{\rm{cen}}=0.52$. Results of the magnitude-limited CMASS sample (faint blue markers and dotted lines) and the corresponding $108$ random catalogs (brown markers) are also shown for comparison. The deviations of the CMASS sample from the $1,000$ random data are indicated in unit of standard deviation ($\sigma$).}
\label{fig:CMASS-mean-std}
\end{figure}

\begin{figure}
\includegraphics[width=\columnwidth,trim=0 20 0 0]{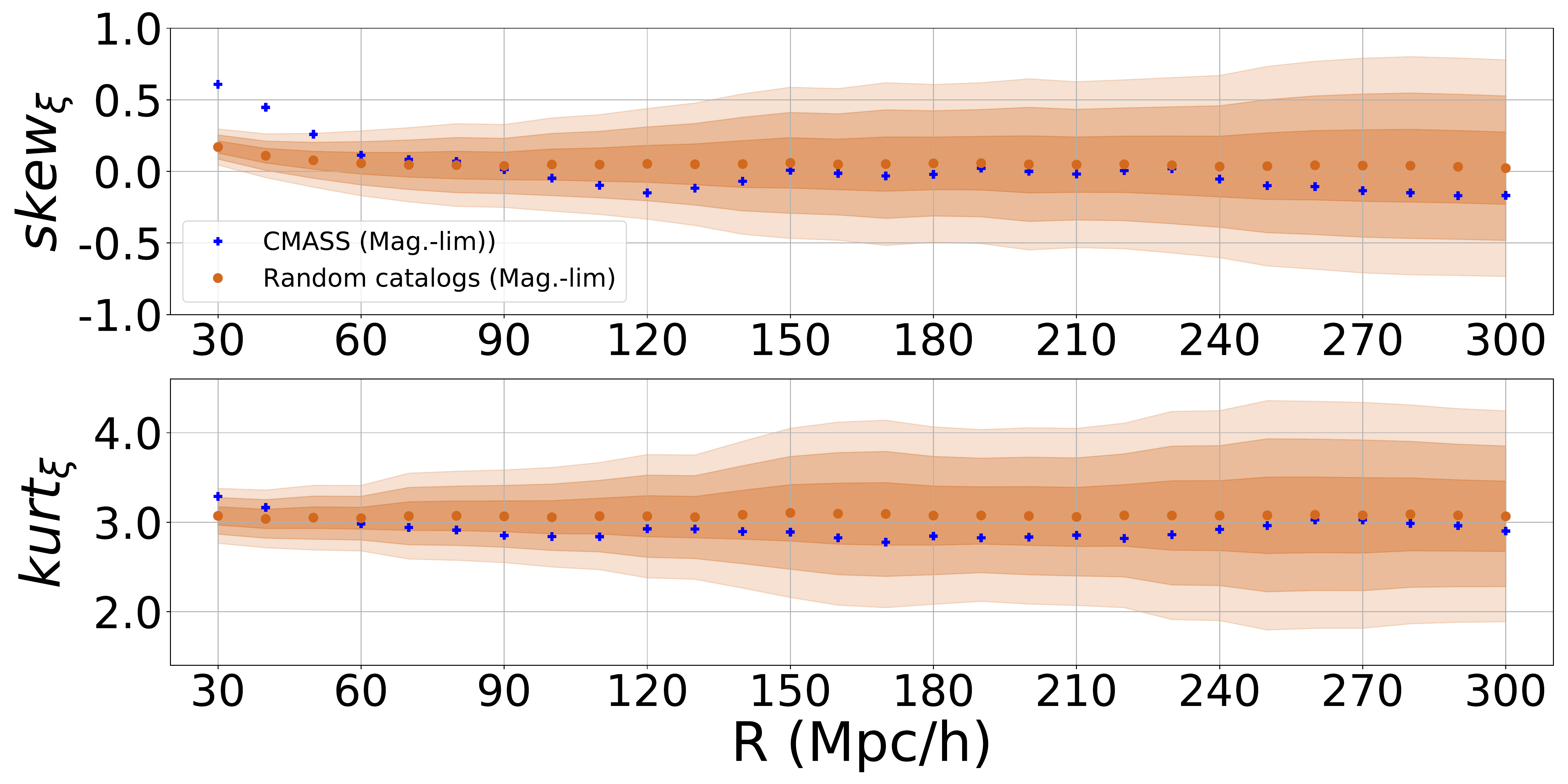}
\caption{The skewness ($skew_\xi$; top) and kurtosis ($kurt_\xi$; bottom) of $\xi$ as a function of radius ($R$). The result of the magnitude-limited CMASS sample (blue dots) is compared with the average and 1--3$\sigma$ confidence regions of the $108$ random catalogs.}
\label{fig:CMASS-skewkurt}
\end{figure}

\begin{figure}
\includegraphics[width=90mm,trim=30 30 20 0]{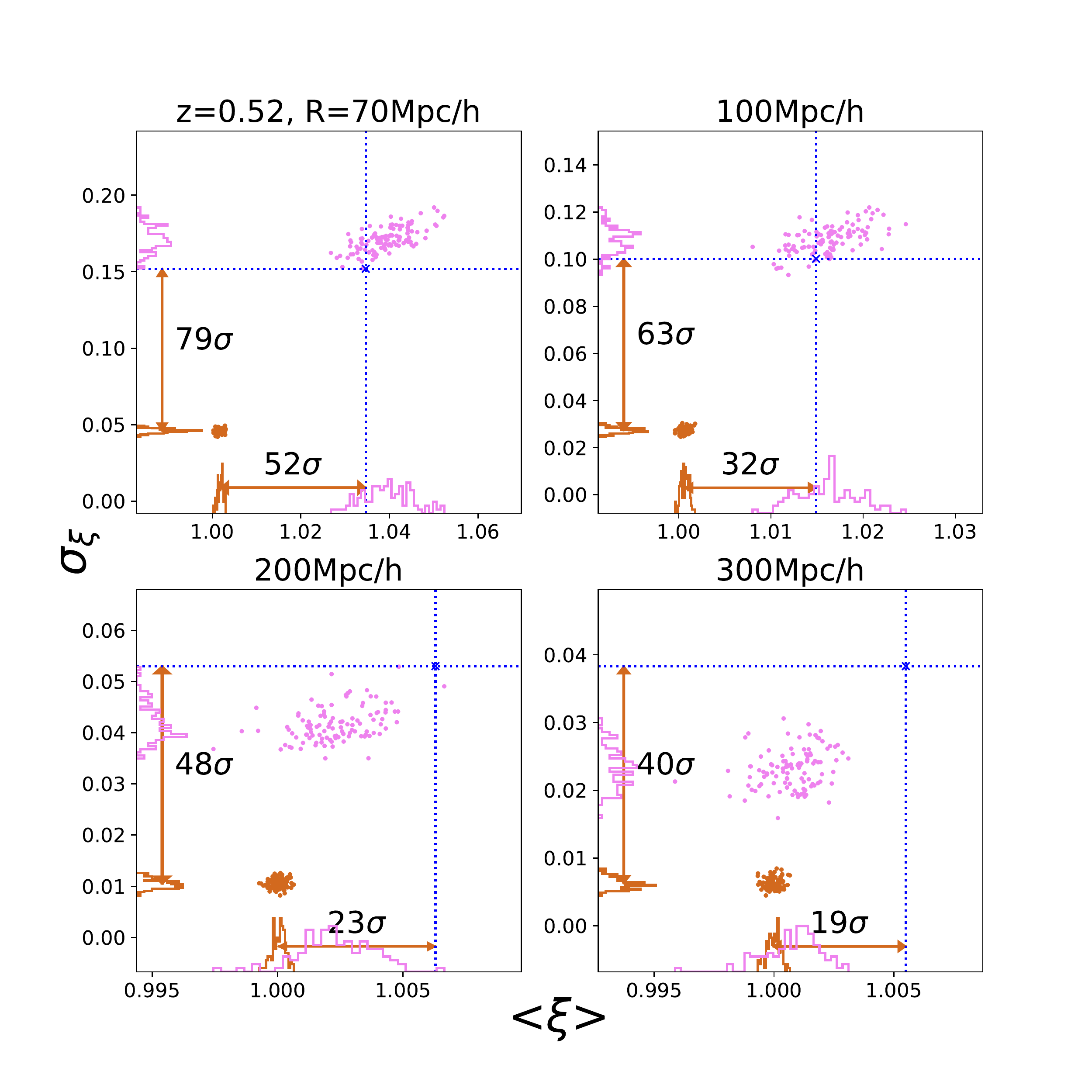}
\caption{The same as Fig.\ \ref{fig:CMASS-mean-std} but for the magnitude-limited CMASS sample, $108$ HR3 mock catalogs (pink dots and histograms), the corresponding $108$ random catalogs (brown dots and histograms).}
\label{fig:CMASS-mean-std-magcut}
\end{figure}

\subsection{Results}\label{sec:result}
\subsubsection{Comparison with random data}

As the homogeneous distribution, we take the random Poisson distribution. Comparisons are made in Figs.\ \ref{fig:CMASS-stat}{a-b}, which show the mean ($\langle\xi\rangle$) and the standard deviation (${\sigma}_{\xi}$) of $\xi$'s calculated from the unmodified CMASS sample and the corresponding $1,000$ random catalogs. We can see that $\langle\xi\rangle$'s of both CMASS sample and random data approach 1 as the radius increases. However, unlike the random distributions, the $\left<\xi\right>$ of the CMASS sample does not fully converge to 1 but shows a large deviation, which maintains even at the largest radius. The $\sigma_\xi$ also significantly deviates from the expected range of the random distributions. For $\langle\xi\rangle$, the deviations are $61\sigma$ at $70~h^{-1}\textrm{Mpc}$ and $5.5\sigma$ at $300~h^{-1}\textrm{Mpc}$; and for ${\sigma}_{\xi}$, the deviations are $93\sigma$ at $70~h^{-1}\textrm{Mpc}$ and $34\sigma$ at $300~h^{-1}\textrm{Mpc}$, see Fig.\ \ref{fig:CMASS-mean-std}. The deviations in the mean and standard deviation show that CMASS galaxies are far from the random distributions even at the largest scale. On the other hand, the skewness and kurtosis of $\xi$-statistics presented in Fig.\ \ref{fig:CMASS-skewkurt} show no significant anomaly of the CMASS sample compared with the random data. Also shown are the trends in the magnitude-limited CMASS sample and random data, where deviations from the random data are far more significant than the original ones.

\subsubsection{Comparison with two mock catalogs}

Figures \ref{fig:CMASS-stat}{c-d} and \ref{fig:CMASS-stat}{e-f} compare $\left<\xi\right>$ and $\sigma_\xi$ variations of the CMASS sample with those of HR3 and log-normal mock catalogs, respectively. Here, the magnitude-limited  CMASS sample and the corresponding random catalogs have been used. For both $\langle\xi\rangle$ and $\sigma_\xi$, the CMASS sample relatively agrees well with both mock samples
(see Fig.\ \ref{fig:CMASS-mean-std-magcut} for HR3 mock samples). We note that the log-normal galaxies generated with a lower value of bias factor ($b=1.8$) have lower amplitudes for the mean and standard deviation of the $\xi$ statistic. Thus, the CMASS galaxies are distributed as expected from theoretically estimated distribution.

\begin{figure}
\includegraphics[width=\columnwidth,trim=0 20 0 0]{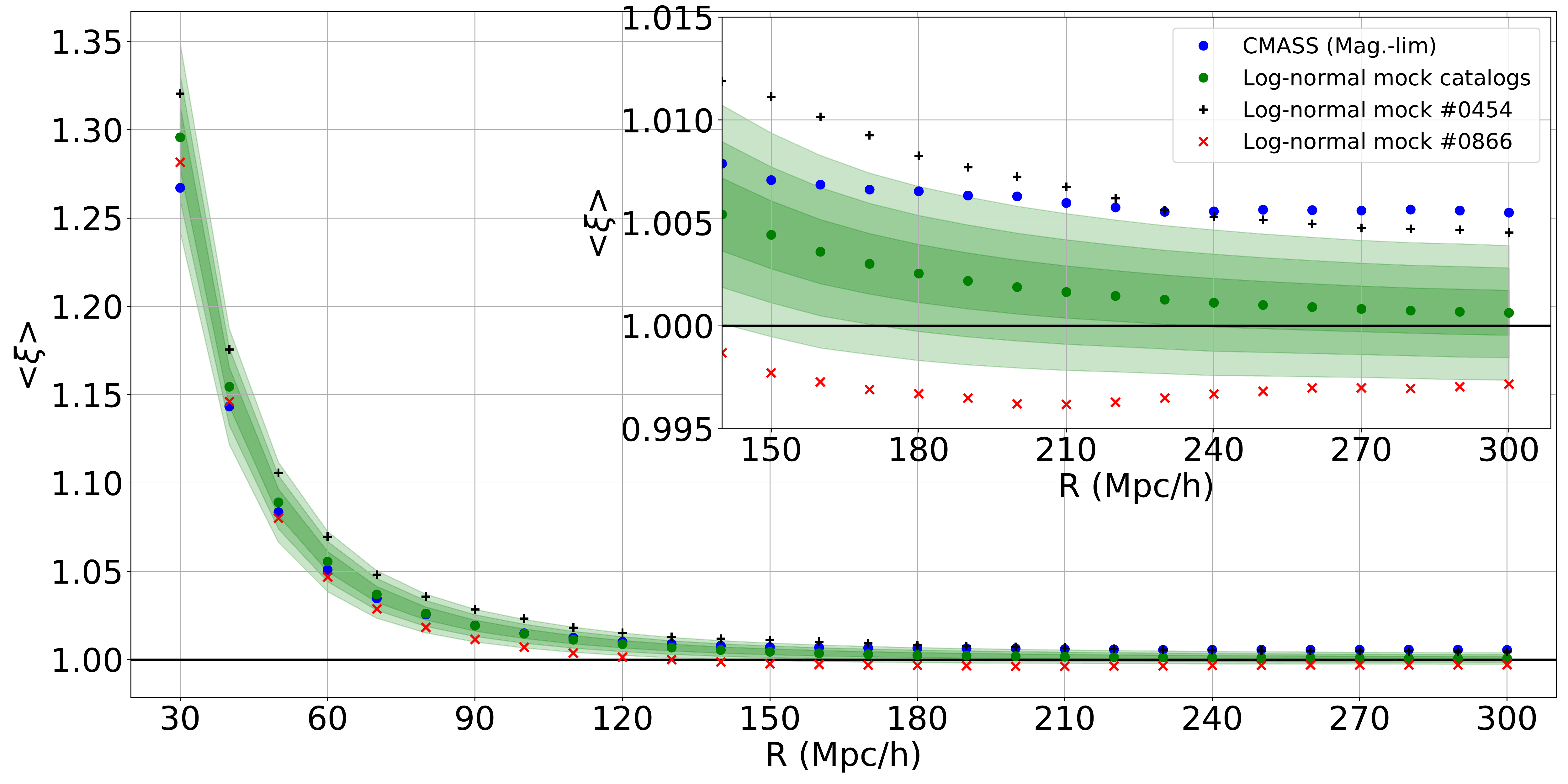}
\caption{Variation of $\left<\xi\right>$ as a function of radius $R$ for log-normal mock catalogs (green dots with shaded regions) and the magnitude-limited CMASS sample (blue dots), which is the same as Fig. \ref{fig:CMASS-stat}{e}. Here the cases of the maximum (black pluses) and the minimum (red cross-marks) $\left<\xi\right>$ at $300~h^{-1}\textrm{Mpc}$ among the 1000 log-normal mock catalogs have been added.}
\label{fig:CMASS-lognormal-extreme-plot}
\end{figure}

\begin{figure}
\includegraphics[width=\columnwidth, trim=0 0 0 0]{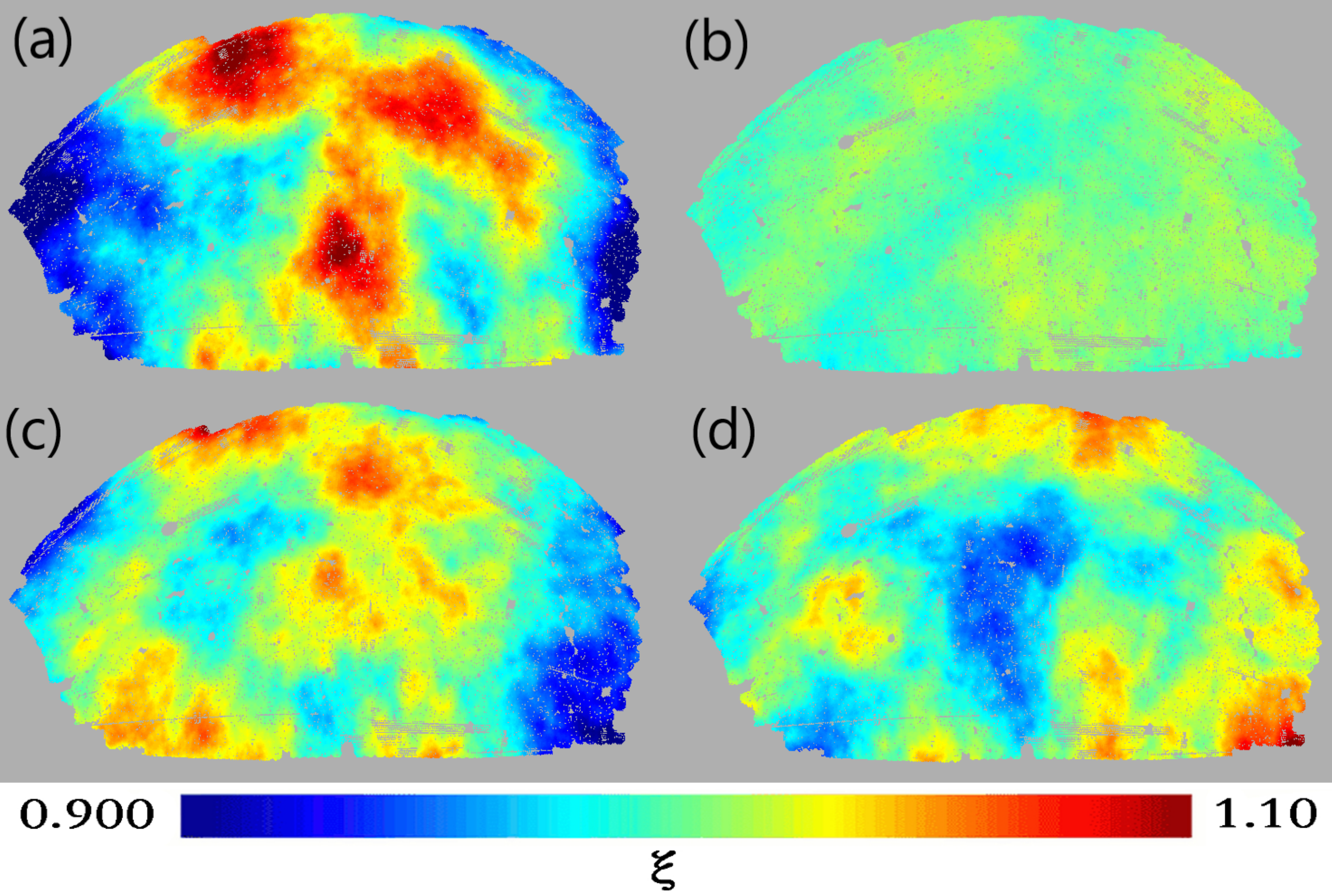}
\caption{Maps showing $\xi$-distributions on the CMASS survey region, estimated from the magnitude-limited CMASS sample (a), a (magnitude-limited) random catalog (b), and two log-normal catalogs with the maximum (c) and minimum (d) values of $\left<\xi\right>$ at $R=300~h^{-1}\textrm{Mpc}$ scale.}
\label{fig:CMASS-lognormal-extreme-map}
\end{figure}

For $\left<\xi\right>$, although the values of mock catalogs tend to approach the result of the random distribution as the scale increases, the $\left<\xi\right>$ of the CMASS sample tends to stay constant starting from $230~h^{-1}\textrm{Mpc}$. This {\it anomaly} can be {\it explained} by the presence of peculiar overdense regions in the CMASS sample. If there are noticeable overdense regions with more clustering within the survey volume, naturally $\langle\xi\rangle$ is expected to be greater than 1. Although not a major effect, it should also be taken into account that galaxies located near the CMASS boundary contribute less to $\left<\xi\right>$ because in this case only the partial volume of the truncated cone is used as weight.
The constant behavior of $\left<\xi\right>$ at large scales with a significant deviation from unity can be also found in the mock catalogs. Figure \ref{fig:CMASS-lognormal-extreme-plot} shows the behavior of $\left<\xi\right>$ as a function of radius scale for the cases of the maximum and minimum $\left<\xi\right>$ at $R=300~h^{-1}\textrm{Mpc}$ among the 1000 log-normal mock catalogs. Figure \ref{fig:CMASS-lognormal-extreme-map} compares the distributions of $\xi$ on the CMASS survey region for the magnitude-limited CMASS sample, a (magnitude-limited) random catalog, and two log-normal mock catalogs that show the maximum or minimum values of $\left<\xi\right>$ at $300~h^{-1}\textrm{Mpc}$ scale (see section \ref{sec:landscape} for a detailed description of how to produce these maps). In the $\xi$ distribution of the CMASS sample, it is seen that the more overdense area exists in the central part of the survey region. In the case of the mock catalog showing the maximum $\left<\xi\right>$ value, the overall distribution is similar. On the other hand, for the mock catalog with the minimum $\left<\xi\right>$, it appears that there are relatively few overdense areas in the inner survey region.

In $\sigma_\xi$, both kinds of mock samples {\it deviate significantly} from the random data, and the deviation becomes even larger for the CMASS galaxies at the largest scale (see Figs. \ref{fig:CMASS-stat}{d}, \ref{fig:CMASS-stat}{f}, and \ref{fig:CMASS-mean-std}). As shown in Fig. \ref{fig:CMASS-mean-std-magcut}, the deviations of the magnitude-limited CMASS galaxies from the random expectations are $19\sigma$ for $\left<\xi\right>$ and $40\sigma$ for $\sigma_\xi$ at $R=300~h^{-1}\textrm{Mpc}$. We note that the CMASS galaxies show similar $\left<\xi\right>$ and $\sigma_\xi$ to mock galaxies on small scales, but the difference between the two becomes more significant as the scale increases. At the largest scale, the CMASS sample tends to deviate by $3\sigma$ or more from the mock samples. Further investigation is needed to explain this discrepancy.

\subsubsection{Comparison with theoretical power spectrum}

In the previous subsection, we have compared the distribution of the CMASS galaxies with those in two theoretical mock catalogs. The distribution of galaxies in the mock catalogs contains information of the nonlinear mass fluctuations. Here we investigate whether the linear matter density fluctuations, which are the result of the primordial fluctuations grown by the linear perturbation theory, can explain the CMASS result without taking into account the effects of nonlinear gravitational evolution. For this, we compare our estimation of $\sigma_\xi$ of the CMASS sample with the theoretically estimated amplitude of mass fluctuations. The variance of mass fluctuations expected within a sphere of radius $R$ is given by \citep{peebles1993}
\begin{equation}
    \sigma^{2}(R) \equiv \langle{\delta({\bf x},R)^{2}}\rangle = {{1}\over{{2\pi}^{2}}}\int P(k)|{W_R({\bf k}})|^{2} k^2 dk,
\label{eq:2}
\end{equation}
where $P(k)$ denotes the matter power spectrum at wavenumber $k$, and $W_R({\bf k})$ is the Fourier transformation of a spherical top-hat window function that assigns $3/(4\pi R^3)$ inside the sphere and $0$ otherwise and is written as
\begin{equation}
    W_R(\mathbf{k})=\frac{3}{(kR)^3} \left[\sin(kR)-kR\cos(kR) \right].
\end{equation}

\begin{figure}
%\begin{center}
\includegraphics[width=\columnwidth,trim=10 0 0 0]{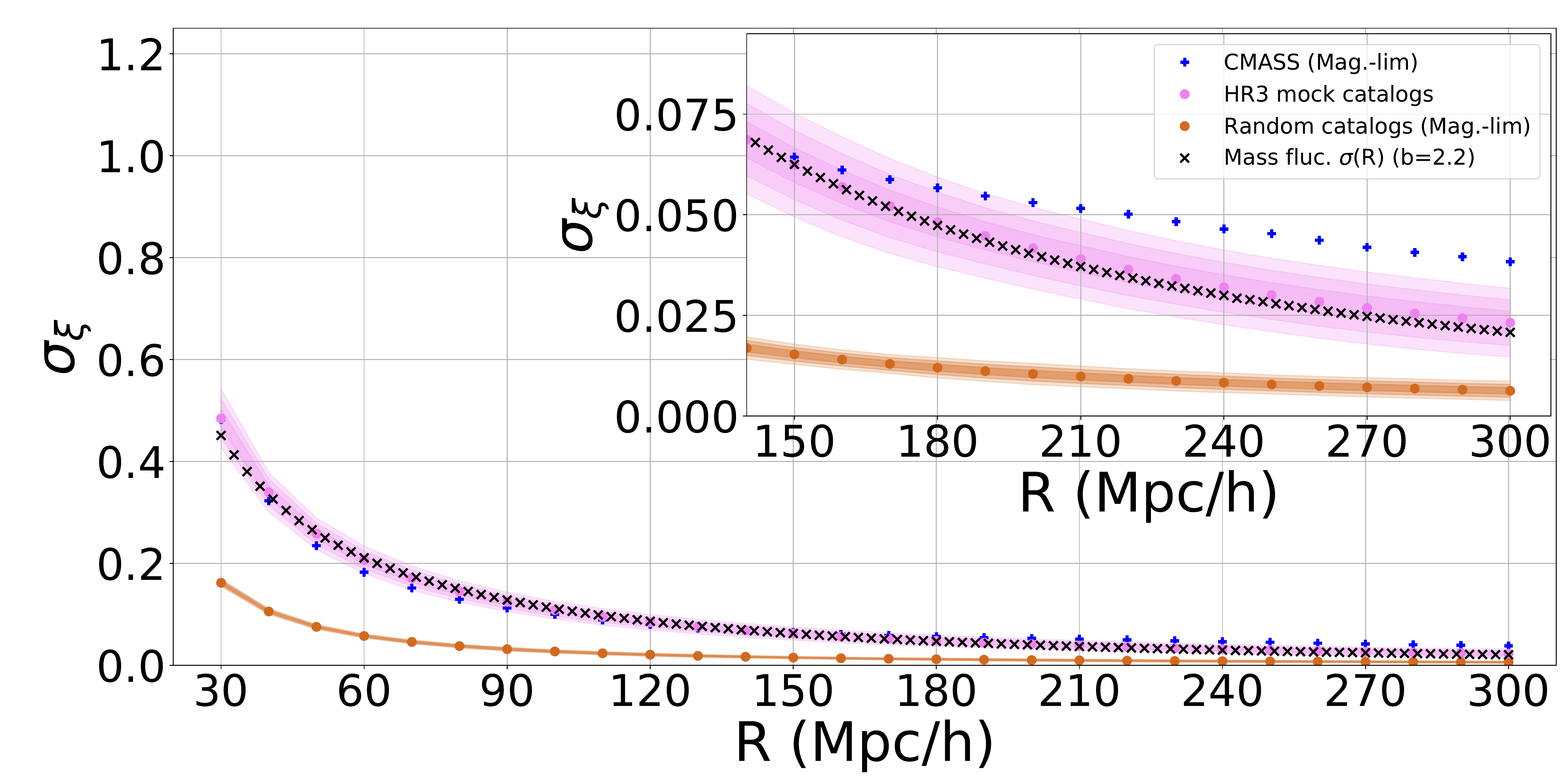}
\includegraphics[width=\columnwidth,trim=10 0 0 0]{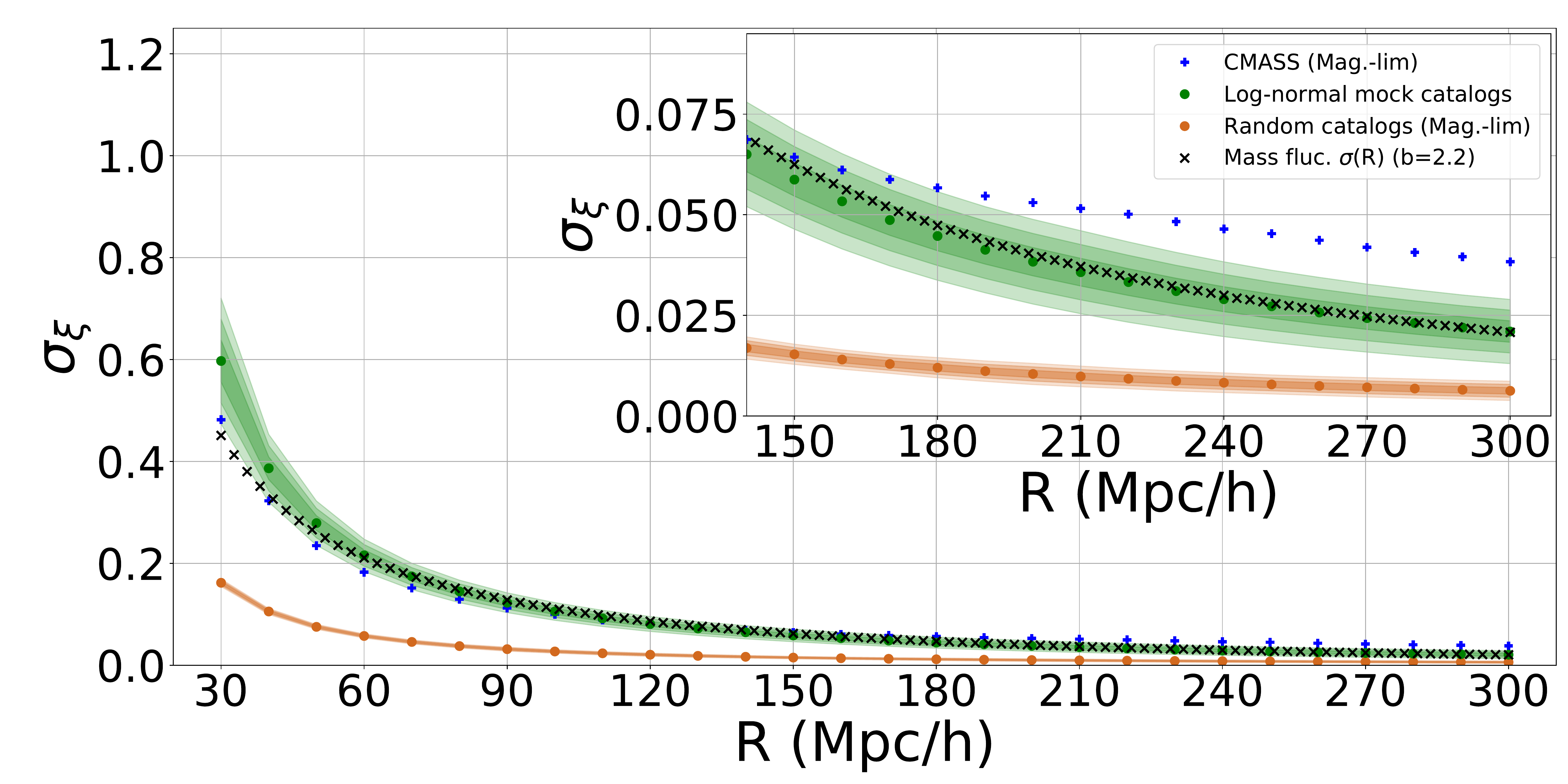}
%\end{center}
%\vspace{-1em}
\caption{Comparison of $\sigma_\xi$ of the CMASS sample, and the HR3 and log-normal mock catalogs (same as Figs.\ \ref{fig:CMASS-stat}{d} and \ref{fig:CMASS-stat}{f}) together with the mass fluctuations $\sigma(R)$ of linear density fields expected in the standard $\Lambda$CDM model \citep{planck2020}.}
\label{fig:CMASS-sigma}
\end{figure}

\begin{figure*}
\begin{center}
\includegraphics[width=160mm,trim=0 30 0 50]{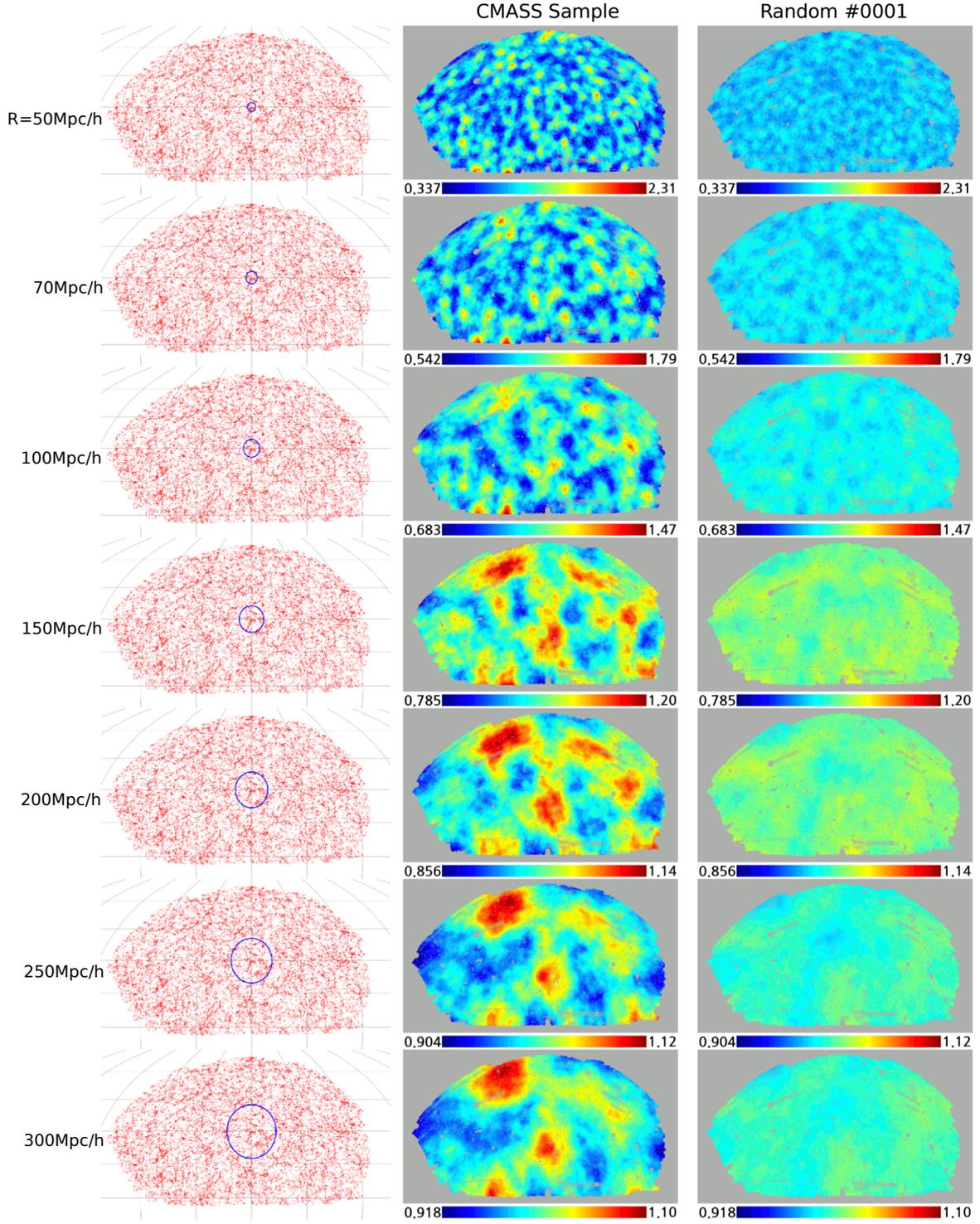}
\end{center}
\caption{Left panel: The CMASS galaxies within a thin slice at the central redshift $z_{\rm{cen}}$. Circles of different sizes of radius ($R=50$, $70$, $100$, $150$, $200$, $250$, $300h^{-1}\textrm{Mpc}$) are shown together. The redshift ranges are presented in Fig.\ \ref{fig:CMASS-radial}. Middle and right panels: Maps of $\xi$ estimated from the unmodified CMASS sample (left) and a random catalog (right panel).}
\label{fig:CMASS-2D}
\end{figure*}

Figure \ref{fig:CMASS-sigma} compares the theoretical prediction of matter density fluctuations $\sigma(R)$ at various scales with the measured deviation of galaxy number density $\sigma_\xi$ of the CMASS sample and two mock catalogs.
In our calculation, we use the CAMB code \citep{camb} to obtain the linear matter power spectrum at $z=0.52$ in a flat $\Lambda$CDM cosmological model that best fits the Planck 2018 TT,TE,EE+lowE data \citep{planck2020}. The $\sigma(R)$ shown in Fig.\ \ref{fig:CMASS-sigma} is obtained by inserting the linear power spectrum into Eq.\ (\ref{eq:2}) and multiplying with the bias factor $b=2.2$, which is consistent with the values estimated in the previous studies \citep{guo2013, cbias2, lbias3}. The $\sigma(R)$ obtained in this way is in good agreement with $\sigma_\xi (R)$ of both HR3 and log-normal mock catalogs at all scales above $50~h^{-1}\textrm{Mpc}$.

%%%%%%%%%%%%%%%%%%%%%%%%%%%%%%%%%%%%%%%%%%%%%%%%%%%%%%%%%%%%%%%
%
%
%
%%%%%%%%%%%%%%%%%%%%%%%%%%%%%%%%%%%%%%%%%%%%%%%%%%%%%%%%%%%%%%%

\subsection{Cosmic landscape}
\label{sec:landscape}

The count-in-truncated cone method used here to test the homogeneity of the galaxy distribution is also useful for visualizing the results and making intuitive comparisons.
Figure \ref{fig:CMASS-2D} visualizes the clustered nature of the CMASS galaxies compared to a random realization. Whereas our statistical studies were based on galaxy-centered counting, the results shown here are based on the pixel-centered counting. We divided the survey area into about $132,000$ pixels using the HEALpix program with a resolution parameter as $\textrm{Nside}=256$ \citep{healpix}. Then, we placed a  truncated cone in the center of every pixel and calculated $\xi$. Figure \ref{fig:CMASS-2D} is the $\xi$-distribution maps and directly shows how the averaged-over-the-radius density changes with position. To compare with the CMASS sample, just one random catalog is selected out of the $1,000$ catalogs.

Figure \ref{fig:CMASS-2D} shows that the CMASS galaxies are far more clustered than the random distribution in all scales, and Table \ref{tab:min_max} compares the density extremum (maximum and minimum) between the CMASS sample and the random data. For the often claimed HS, $70~h^{-1}\textrm{Mpc}$, the density maximum of the CMASS sample is $79\%$ higher than the mean, while for the random one, it is only $18\%$; the ratio between the maximum and minimum is $3.30$ for the CMASS sample, while it is only $1.39$ for the random data. It is not easy to call a distribution with such large fluctuations homogeneous.
Even at $300~h^{-1}\textrm{Mpc}$, it is difficult to say that the distribution of galaxies is homogeneous: $\xi$'s at maximum for the CMASS and random data are $10\%$ and $2\%$ higher than mean, respectively, and the ratio between the maximum and minimum is $1.20$ versus $1.05$.

\begin{table*}
\footnotesize
	\centering
	\begin{tabular}{cccccccc}
		\hline\\[-2mm]
		 & $50~h^{-1}\textrm{Mpc}$ & $70~h^{-1}\textrm{Mpc}$ & $100~h^{-1}\textrm{Mpc}$ & $150~h^{-1}\textrm{Mpc}$ & $200~h^{-1}\textrm{Mpc}$ & $250~h^{-1}\textrm{Mpc}$ & $300~h^{-1}\textrm{Mpc}$ \\
		\hline\\[-2mm]
		CMASS galaxies & 33.7 $\sim$ 231\% & 54.2 $\sim$ 179\% & 68.3 $\sim$ 147\% & 78.5 $\sim$ 120\% & 85.6 $\sim$ 114\% & 90.4 $\sim$ 112\% & 91.8 $\sim$ 110\% \\
		Random distribution & 71.4 $\sim$ 133\% & 84.6 $\sim$ 118\% & 89.9 $\sim$ 109\% & 95.4 $\sim$ 105\% & 95.8 $\sim$ 104\% & 97.2 $\sim$ 103\% & 97.5 $\sim$ 102\% \\
		\hline\\[-2mm]
	\end{tabular}
    \caption{Minimum and maximum $\xi$ values for different radius scales in the $\xi$-maps shown in Fig.\ \ref{fig:CMASS-2D}. The values represent deviations from the mean (100\%) of the maps.}
	\label{tab:min_max}
\end{table*}

%%%%%%%%%%%%%%%%%%%%%%%%%%%%%%%%%%%%%%%%%%%%%%%%%%%%%%%%%%%%%%%
%
%
%
%%%%%%%%%%%%%%%%%%%%%%%%%%%%%%%%%%%%%%%%%%%%%%%%%%%%%%%%%%%%%%%
\subsection{Conclusion}\label{sec:conclusion}

In this section, we studied the inhomogeneous nature of galaxy distribution using the BOSS CMASS DR12 sample, which includes $932,517$ galaxies within the redshift range between $0.43$ and $0.7$ \citep{alam2015}. We use the number count within truncated cones centered at every galaxy located at a narrow region around redshift $0.52$. As the criterion of homogeneity, we statistically compared with the $1,000$ realization of random Poisson distributions. Statistical comparisons are also made with two mock catalogs derived from theory: (i) $108$ mock galaxy catalogs from HR3 $N$-body numerical simulation \citep{horizonrun2011} and (ii) the mock catalogs based on the log-normal distribution of matter density \citep{lognormal2017}. We also compared $\sigma(R)$ estimated from the linear matter power spectrum.

The result shows that the difference between the CMASS sample and the random data reduces as the scale becomes larger, see Figs.\ \ref{fig:CMASS-stat}{a-b}, and \ref{fig:CMASS-skewkurt}. However, even in the largest statistically meaningful scale, the radius around $300~h^{-1}\textrm{Mpc}$, the difference is still significant. Even at the largest scale, the deviation of the CMASS sample from the Poisson distribution is around $34\sigma$ to $40\sigma$ in standard deviation ($\sigma_\xi$) and around $5\sigma$ to $19\sigma$ in the mean ($\left< \xi \right>$), see Figs.\ \ref{fig:CMASS-mean-std} and \ref{fig:CMASS-mean-std-magcut}. Thus, even on this largest available scale, the CMASS data is impossible to appear in random realizations. The cosmic landscape shown in Fig.\ \ref{fig:CMASS-2D} visually confirms our conclusion. The CMASS galaxies are far more clustered than the randomly distributed points in all currently available scales.

However, our comparison of the CMASS sample with the two mock catalogs shows consistency between the observation (CMASS sample) and theory (two mock catalogs), see Figs.\ \ref{fig:CMASS-stat}{c-f}, \ref{fig:CMASS-mean-std-magcut}, and \ref{fig:CMASS-sigma}. A statistical anomaly in $\left< \xi \right>$  identified at a large scale is explained by the presence of overdense regions located in the CMASS survey area. We confirmed the same phenomenon occurs in the log-normal mock catalogs with similar overdensities, see Section \ref{sec:result} and Figs.\ \ref{fig:CMASS-lognormal-extreme-plot} and \ref{fig:CMASS-lognormal-extreme-map}. A statistical anomaly in $\sigma_\xi$ compared with the two mock data identified at a large scale in Figs.\ \ref{fig:CMASS-stat}d, \ref{fig:CMASS-stat}f, and 9 needs further study.

Therefore, we conclude that while the observation is consistent with the current paradigm of modern cosmology, the CP is not in the currently available galaxy distribution sample and nor is it demanded by theory. A clear resolution of this seemingly contradictory conclusion is made in Sections \ref{sec:homogeneity} and \ref{sec:discussion}.

%%%%%%%%%%%%%%%%%%%%%%%%%%%%%%%%%%%%%%%%%%%%%%%%%%%%%%%%%%%%%%%
%
%
%
%%%%%%%%%%%%%%%%%%%%%%%%%%%%%%%%%%%%%%%%%%%%%%%%%%%%%%%%%%%%%%%
\section{Discussion}\label{sec:discussion}

Einstein's gravity directly connects the matter with the spacetime curvature. However, the CP is implemented in the metric. As the curvature is a quadratic derivative of the metric, the near-homogeneity in the metric does not demand similar near-homogeneity in the matter distribution. Such a difference is indeed the case in modern cosmology, especially on the far sub-horizon scale. Our study in Section \ref{sec:test} confirms that matter distribution is far more inhomogeneous compared with the random Poisson distribution, which is our criterion for the homogeneity. At the same time, we show that the matter distribution is consistent with theoretical predictions based on simulation and power spectrum. Thus, the CP is not in the observed galaxy distribution, nor is it expected in theory.

We conclude that it is {\it inadequate} to search for the HS or the CP in the observed light or the matter distribution. As we show in Section \ref{sec:homogeneity} the CP is perfectly fine and well {\it in} the metric, a theoretical construct. However, this conclusion is valid under a subtle condition. Based on the Robertson-Walker metric as the background, we have shown that the deviations in {\it metric} are extremely small in the observed cosmic scale. Thus, our test implies that the CP is valid in modern cosmology {\it if} we assume the CP in the background. That is, our's is only a {\it consistency} test, not proof. Whether we can accept such a background geometry is an entirely different question that may demand another strategy. Until we resolve this aspect, the CP remains an assumption.

A proper answer to this question may require studying alternative models based on other than Einstein's CP. The CMB strongly supports the isotropy around us. Thus, one interesting possibility is constraining the spherically symmetric model based on Lema\^itre's CP. After all, due to the nature of our observation, systematically observing past events as we observe further away, even under Einstein's CP properly unfolded in the Universe, the observations are destined to be spherically symmetric with us at the center. The temporal evolution of cosmic structures and the observational difficulties increasing with distance will make the observation look spherically symmetric in any case. Constraining the spherically symmetric world models with us near the center is an important issue in this regard.

One consequence of our clarification, separating matter from the metric, is freeing our concept of the universe from dull homogeneity in its {\it matter} content on the large-scale because the CP in modern cosmology mainly concerns the metric only. Through the light, one may discover more diverse phenomena that may have distinct characteristics and may await new classification. Perhaps, if we may, ``there are more things in the sky than are dreamt of in our theory.''

Although we have not found the HS in the currently available galaxy survey, this does not preclude its future discovery in a more in-depth survey. In modern cosmology, with $\delta \Phi/c^2 \sim (\ell/\ell_H)^2 \delta$, as the scale approaches the horizon scale, we should have $\delta \sim \delta \Phi/c^2$, thus the matter fluctuation ($\delta$) becoming extremely small as the metric fluctuation. In this way, the Friedmann background world model should be recovered with the CP valid in both metric and matter, in near-horizon scale structures. No such confirmation has been made yet. We anticipate practical difficulty as the scale approaches the horizon scale; the observed space (light-cone) deviates significantly from the homogeneous space the CP is imposed. Still, although not relevant to the CP's validity, the homogeneity test for matter should be continued using more extensive galaxy survey data undoubtedly available soon in the future.

Now, with the extreme level homogeneity in the cosmic {\it metric} known, an important theoretical issue arises concerning its origin. Inflation provides an isotropization mechanism for homogeneous but anisotropic cosmologies \citep{wald1983}, but how about homogeneity? Why do we live in a universe with such extreme homogeneity in the metric? This is a mystery yet to be resolved. The anthropic principle is not an answer we anticipate as it merely claims the case as it happens to be.

%%%%%%%%%%%%%%%%%%%%%%%%%%%%%%%%%%%%%%%%%%%%%%%%%%%%%%%%%%%%%%%
%
% Acknowledgments
%
%%%%%%%%%%%%%%%%%%%%%%%%%%%%%%%%%%%%%%%%%%%%%%%%%%%%%%%%%%%%%%%
\section*{Acknowledgements}

We wish to thank an anonymous referee for valuable comments and suggestions which help improve the manuscript.
We thank Professors Donghui Jeong and Juhan Kim for useful discussion and suggestions.
C.-G.P. was supported by the National Research Foundation of Korea (NRF) grant funded by the Korea government (MSIT) (No.\ 2020R1F1A1069250).
H.N.\ was supported by the National Research Foundation of Korea funded by the Korean Government (No.\ 2018R1A2B6002466 and No.\ 2021R1F1A1045515).
J.H.\ was supported by IBS under the project code IBSR018-D1 and by Basic Science Research Program through the National Research Foundation (NRF) of Korea funded by the Ministry of Science, ICT and Future Planning (No.\ 2018R1A6A1A06024970 and NRF-2019R1A2C1003031).

%%%%%%%%%%%%%%%%%%%%%%%%%%%%%%%%%%%%%%%%%%%%%%%%%%
%\section*{Data Availability}
%
%All results presented in this article were derived from the publicly available data. The BOSS DR12 CMASS galaxy sample is available on the website of the Sloan Digital Sky Survey Data Release 12 (\url{https://www.sdss.org/dr12/data_access/}. The mock SDSS-III survey data are available on the website of The Horizon Runs (\url{http://sdss.kias.re.kr/astro/Horizon-Runs/}). The code for generating the log-normal realizations of galaxies is available on the website \url{https://bitbucket.org/komatsu5147/lognormal_galaxies/src/master/}. We also acknowledge the use of the HEALpix, MANGLE, and CAMB codes.

%\bibliography{sample631}{}
%\bibliographystyle{aasjournal}



\def\and{{and }}
\bibliographystyle{yahapj}

\end{document}